 \documentclass[a4paper,twoside,11pt]{article}

\usepackage{amsmath}
\usepackage{amsfonts}
\usepackage{natbib}
\usepackage{hyperref}
\usepackage{amssymb}
\usepackage[british]{babel}
\usepackage[sans]{dsfont}
\usepackage{subfigure}
\usepackage{wrapfig}
\usepackage{amsthm}
\usepackage{latexsym}
\usepackage{mathrsfs}
\usepackage[mathscr]{euscript}
\usepackage{array}
\usepackage{verbatim}
\usepackage[left=2.5cm,top=3cm,right=2.5cm,nohead,foot=1cm]{geometry}
\usepackage{placeins}
\usepackage{vruler}
\usepackage{color}

\usepackage{graphicx}
\usepackage{epstopdf}
\graphicspath{{Figures/}}
\usepackage[font=scriptsize]{caption}

\newtheorem{theorem}{Theorem}[section]
\newtheorem{lem}[theorem]{Lemma}
\newtheorem{pro}[theorem]{Proposition}
\newtheorem{cor}[theorem]{Corollary}
\newtheorem{conj}[theorem]{Conjecture}

\newtheorem{rem}[theorem]{Remark}
\newtheorem{com}[theorem]{Comments}
\newtheorem{ex}[theorem]{Example}
\newtheorem{defi}[theorem]{Definition}
\newtheorem{hyp}[theorem]{Assumption}

\newcommand{\bt}{\begin{theorem}}\newcommand{\et}{\end{theorem}}
\newcommand{\bl}{\begin{lem}}\newcommand{\el}{\end{lem}}
\newcommand{\bp}{\begin{pro}}\newcommand{\ep}{\end{pro}}
\newcommand{\bcor}{\begin{cor}}\newcommand{\ecor}{\end{cor}}
\newcommand{\bconj}{\begin{conj}}\newcommand{\econj}{\end{conj}}

\newcommand{\bd}{\begin{defi} \rm }\newcommand{\ed}{\end{defi} }
\newcommand{\brem }{\begin{rem} \rm }\newcommand{\erem }{\end{rem}}
\newcommand{\bcom}{\begin{com} \rm }\newcommand{\ecom }{\end{com}}
\newcommand{\brems }{\begin{rem} \rm }\newcommand{\erems }{\end{rem}}
\newcommand{\bex}{\begin{ex} \rm }\newcommand{\eex}{\end{ex}}
\newcommand{\bhyp}{\begin{hyp} \rm }\newcommand{\ehyp}{\end{hyp}}

\def\proof{\noindent \textbf{\emph{\textbf{Proof}.$\,$}}}
\def\finproof {\hfill $\Box$ \vskip 5 pt }

\def \be{\begin{eqnarray}}
\def \ee{\end{eqnarray}}
\def \b*{\begin{eqnarray*}}
\def \e*{\end{eqnarray*}}

\def\Q{\mathbb Q}

\def \[{[\,\!\![}
\def \]{]\,\!\!]}

\def \1{{\bf 1}}

\def \proof{{\noindent \bf Proof. }}

\def\F{{\cal T}}\def\F{{\cal G}}\def\F{{\cal F}}

\def\E{\mathbb{E}}

\def\F{{\cal F}}

\newcommand{\bea}{\begin{eqnarray*}}
\newcommand{\eea}{\end{eqnarray*}}
\newcommand{\beqa}{\begin{eqnarray}}
\newcommand{\eeqa}{\end{eqnarray}}

\def\proof{\noindent {\it Proof. $\, $}}
\def\finproof {\hfill $\Box$ \vskip 5 pt }

\def\label{\label}

\def\bal{\begin{aligned}}
\def\eal{\end{aligned}}

\def\thetau{\tau}

\def\xitau2{\xi_{(\thetau)}}
\def\xiitau2{\xi^i_{(\thetau)}}
\def\xintau2{\tilde{\xi}_{(\thetau)}}
\def\Ltau2{\tilde{\xi}(\tau_0,\tau_1)}
\def\chiitau2{P^i_{\tau}}

\def\emph{}

\def\imath{Y}

\def\varkappa{\mathbf{l}}

\def\eee{\end{document}}

\newcommand{\beql}[1]{\beqa\label{#1}\begin{aligned}}

\newcommand{\eel}{\end{aligned}\eea}

\begin{document}

\title{{\LARGE
On the Range of Admissible Term Structures
\vspace{9pt}
}}

\small{
\author{
 Areski Cousin\thanks{The research of A. Cousin benefited from the support of the ``chaire d'excellence management de la mod\'elisation''. The authors are grateful to V\'eronique Maume-Deschamps for many helpful discussions on the subject.}~,\, 
Ibrahima Niang
\vspace{5pt}\\\\
Universit\'e de Lyon,\\
 Universit\'e Lyon 1, Laboratoire SAF, EA 2429, \\
 Institut de Science Financi{\`e}re et d'Assurances, \\
 50 Avenue Tony Garnier, 69007 Lyon, France.\\\\
}

\maketitle

\begin{abstract}

In this paper, we analyze the diversity of term structure functions (e.g., yield curves, swap curves, credit curves) constructed in a process which complies with some admissible properties: arbitrage-freeness, ability to fit market quotes 
and a certain degree of smoothness. 
When present values of building instruments are expressed as linear combinations of some primary quantities such as zero-coupon bonds, discount factor, or survival probabilities, arbitrage-free bounds can be derived for those quantities at the most liquid maturities. As a matter of example, we present an iterative procedure that allows to compute model-free bounds for OIS-implied discount rates and CDS-implied default probabilities. We then show how mean-reverting term structure models can be used as generators of admissible curves. This framework is based on a particular specification of the mean-reverting level which allows to perfectly reproduce market quotes of standard vanilla interest-rate and default-risky securities while preserving a certain degree of smoothness.
The numerical results suggest that, for both OIS discounting curves and CDS credit curves, the operational task of term-structure construction may be associated with a significant degree of uncertainty.\\

\noindent


\end{abstract}




\section{Introduction}\label{s:intro}

Building financial curves from market quotes of liquidly traded products
 is at the heart of modern asset pricing and risk management. A term-structure curve  describes the evolution of a particular economic or financial variable (such as interest rate, yield-to-maturity, credit spread, volatility) as a function of time-to-maturity.
In most situations, market quotes are only reliable for a small set of liquid instruments whose value may depend on several points of the curve. If all maturities were liquidly traded, then the curve could be unambiguously inferred from market quotes.
Consequently, the financial industry has to rely on somehow arbitrary interpolation or calibration schemes to construct term structure curves. These techniques are used to supplement the market information in illiquid parts of the curve while extracting the quantities of interest for all required maturities.\\

Interestingly, a pretty large recent literature  is devoted to the subject of curve construction methods.  \citet{Hagan2006} provide a review of different interpolation techniques for curve construction. They introduce a monotone convex method and postulate a series of quality criterion such as ability to fit market quotes, arbitrage-freeness, smoothness, locality of interpolation scheme, stability of forward rate and consistency of hedging strategies.
 \citet{Andersen2007} analyzes the use of tension splines for construction of bond term structures. In this approach, stability of forward rates and market fit precision seems to be difficult to achieve simultaneously. 
 \citet{Iwashita2013} makes a survey of non-local spline interpolation techniques which preserve stability of forward rates. 
 \citet{LeFloch2013} introduces another quality criteria related to consistency of hedging strategies. He postulates that, given a constructed term structure, the sum of sequential deltas should be close enough to the corresponding parallel delta. He observes that most of spline techniques are not able to achieve this property correctly.
Other papers such as  \citet{Ametrano2009}, \citet{Chibane2009}, \citet{Kenyon2012} or \citet{Fries2013} are concerned with the adaptation of curve construction methods in a multi-curve interest-rate environment. 
Note that, in terms of interpolation scheme, there is no consensus towards a particular best practice method in all circumstances.\\

In another hand, the question of model uncertainty and its impact on the assessment of financial derivatives has been studied since a certain time period and, following the recent fincancial crisis, has received a particular interest. Impact of model risk on valuation and hedging of financial derivatives have been treated by, among others,  \citet{Derman1996}, \citet{Eberlein1997}, \citet{ElKaroui1998}, \citet{Green1999}, \citet{Branger2004}, \citet{Cont2006}, \citet{Davis2007}, \citet{Henaff2010}, \citet{Morini2011}. In most papers, the question of model risk is restricted to the class of  derivative products and is considered in a pretty theoretical way. In contrast, the question of model risk embedded in the construction process of term-structures  has not been investigated as a main object, whatever it  may concern discount curves, zero-coupon curves, swap basis curves, bond term structures or CDS-implied survival curves. \\

We define admissible curves as arbitrage-free term structures which are perfectly compatible with input market quotes and which admit a minimum degree of smoothness. The aim of this paper is to provide a methodology for assessing the range of these admissible curves and its impact on asset valuation and hedging.
Present values of instruments involved in the curve construction process are assumed to have linear representations in terms of some elementary quantities which define the curve. Depending on the context, these elementary quantities can be discount factors, zero-coupon prices or survival probabilities. Under this assumption and using a no arbitrage argument, we present an iterative procedure which allows to compute model-free bounds for these elementary quantities. We show that admissible curves can be easily generated using mean-reverting term-structure models, which are classically  employed for valuation and hedging of financial derivatives. Contrary to an HJM approach where an arbitrary initial term-structure is given as a model input, the initial curve is here constructed as a by-product of a calibrated term-structure model. The proposed curve construction method relies on a piecewise-constant specification of the mean-reverting level. We identify under which condition the implied parameters lead to an admissible curve.\\

The paper is organized as follows. Section \ref{sec:admissible_curves} defines what is understood as admissible term-structures, with a focus on curves constructed from market quotes of interest-rate sensitive (IR curves) and default-sensitive products (credit curves). Our framework relies on the assumption that present values of building instruments have linear representations with respect to some elementary quantities. We show that this assumption is usually satisfied for construction of bond term-structures, OIS discount curves, forward rate curves implied from market quotes of swaps versus Libor or Euribor index rates. 
  Section \ref{sec:bounds} explains how to compute arbitrage-free bounds for OIS discount curves and CDS survival curves.
In Section \ref{sec:construct_admissible_curve}, we develop a methodology that allows to generate admissible curves as a by-product of a mean-reverting term-structure model. By playing with some free parameters, we illustrate the range of admissible curves that can be obtained in different approaches, for OIS discount curves and CDS-implied survival curves.  Section \ref{sec:conclusion} concludes.

\section{Admissible curves}
\label{sec:admissible_curves}

In this section, we define what is understood as an admissible curve. The presentation of admissible curve is generic in the sense that it covers both term-structure curves constructed from interest-rate or fixed-income products (IR curves) but also default distribution functions constructed from a term-structure of CDS spreads (Credit curves). 
Let us first define what is an arbitrage-free curve in these two cases.

\begin{defi}[arbitrage-free curve]
\label{def:arbitragefree}
A curve is said to be arbitrage-free if
\begin{itemize}
	\item \textbf{IR curves} : the associated forward rates are non-negative or equivalently, the associated zero-coupon prices are nonincreasing with respect to time-to-maturities.
	\item \textbf{Credit curves} : the curve corresponds to a well-defined default distribution function.
\end{itemize}
\end{defi}

For IR curves, the no-arbitrage property means that forward rates associated with any future loan period should be positive. 
In addition, admissible curves are also required to be smooth at a certain minimal degree.

\begin{defi}[Smoothness condition]
\label{def:smooth}
A curve is said to be smooth if 
\begin{itemize}
	\item \textbf{IR curves} : the associated instantaneous forward rates exist for all maturities and are continuous.
	\item \textbf{Credit curves} : the associated default density function exists and is continuous.
\end{itemize}
\end{defi}

As pointed out  in \citet{McCulloch2000}, ``a discontinuous forward curve implies either implausible expectations about future short-term interest rates or implausible expectations about holding period returns''.


\begin{defi}[Admissible curve]
\label{Def:admissible}
Given a set of observed market quotes. A curve is said to be admissible if it satisfies the  following three constraints:
\begin{itemize}
\item The selected market quotes are perfectly reproduced by the curve. 
\item The curve is arbitrage-free in the sense of Definition \ref{def:arbitragefree}.
\item The curve satisfies the smoothness condition presented in Definition \ref{def:smooth}.
\end{itemize}
\end{defi}

Term-structure functions are usually constructed from market quotes of a small number of liquidly traded instruments. We assume that the present values of these market instruments can be represented as linear combinations of some elementary quantities. As illustrated in the examples below, these elementary quantities can be  OIS discount factors, Libor or Euribor forward rates or CDS-implied survival probabilities.
 
 \begin{hyp}[Linear representation of present values]
 \label{Ass:linear}
Present values of products used in the curve construction process can be expressed as linear combination of some elementary quantities. Depending on the context, the latter can be either zero-coupon prices, discount factors, Libor or Euribor forward rates or CDS-implied survival probabilities.
\end{hyp}
 


Let us now present some situations where Assumption \ref{Ass:linear} holds. In what follows, $t_0$  denotes the quotation date. 


\begin{ex}[Corporate or sovereign bond yield curve] 
\label{ex:yield_curve}
Let $S$ be the observed market price  of a corporate or a sovereign bond with maturity time $T$ and with a fixed coupon rate $c$. The price $S$ and the coupon rate $c$ are expressed in percentage of invested nominal. The set of coupon payment dates is given by $(t_1,\ldots, t_p)$ where $t_0<t_1<\ldots<t_p=T$ and $\delta_k$ represents the year fraction corresponding to period $(t_{k-1}, t_k)$, $
k=1, \ldots, p$. The present value of this bond can be defined as a linear combination of some default-free zero-coupon bonds, i.e., 
\begin{equation}
\label{eq:bond}
 c\sum_{k=1}^{p}{\delta_k P^{B}(t_0,t_k)} + P^{B}(t_0, T) = S
\end{equation}
where $P^{B}(t_0,t)$ represents the price at time $t_0$ of a default-free zero-coupon bond with maturity $t$. Note that, even if representation \ref{eq:bond} obviously relies on a default-free assumption, it is commonly used as an intermediary step in the computation of the so-called bond yield-to-maturity.\footnote{The bond yield for maturity time $T$ is defined as the constant rate of return $Y$ such that $P^{B}(t_0,t)$ has a return rate $Y$ for any time $t$, $t_0 \leq t \leq T$ and such that present value relation \ref{eq:bond} holds.}.\\
\end{ex}

%
%

\begin{ex}[Discounting curve based on OIS] 
\label{ex:discount_curve}
Due to legal terms of standard collateral agreements, a possible choice to build discounting curves is to use quotations of OIS-like instruments (See, for instance \citet{Hull2013} for more details).
Let $S$ be the par swap rate of an overnight indexed swap with maturity $T$ and fixed leg payment dates $t_1<\ldots<t_p = T$. If for any $k=1, \ldots, p$, $\delta_k$ represents the year fraction corresponding to period $(t_{k-1}, t_k)$, 
the swap equilibrium relation can be expressed in linear form with respect to some discount factors $P^{D}$ as
\begin{equation}
\label{eq:OIS}
S \sum_{k=1}^{p}  \delta_k P^{D}(t_0,t_k)  = 1 - P^{D}(t_0, T) 
\end{equation}
where $P^{D}(t_0,t)$ is the discount factor associated with maturity date $t$. In the previous equation, the left hand side represents the fixed leg present value whereas  the right hand side corresponds to the floating leg present value. For more details on the derivation of (\ref{eq:OIS}), the reader is referred to \citet{Fujii2010}.
\end{ex}

%
%

\begin{ex}[Forward curve based on fixed-vs-Ibor-floating IRS] 

 Let $S$ be the observed par rate of an interest rate swap with maturity time $T$ and floating payments linked to a Libor or an Euribor rate associated with a tenor $j$ (typically, $j=3$ months or $j=6$ months). The fixed-leg payment scheme is given by $t_1<\cdots<t_p=T$ and  the floating-leg payment scheme is given by $\tilde{t}_1<\cdots<\tilde{t}_{q}=T$. 
  For most liquid products, payment on the fixed  leg are made with an annual frequency, so that $t_k$ is the business day corresponding to $k$ years after the current date $t_0$. The associated year fraction for the interval $(t_{k-1}, t_k)$ is denoted by $\delta_k$ whereas the year fraction for the interval $(\tilde{t}_{i-1}, \tilde{t}_{i})$ is denoted by $\tilde{\delta}_i$. Note that the length between two consecutive dates on the floating leg should be close to the Libor or Euribor tenor, i.e., $\tilde{\delta}_i \simeq j$. As a result, given a discounting curve $P^{D}$, the swap equilibrium relation can be represented in a linear form with respect to some forward Libor or Euribor rates, i.e.,  
\begin{equation}
\label{eq:IRS}
S \sum_{k=1}^{p}  \delta_k P^{D}(t_0,t_k) = \sum_{i=1}^{q} P^{D}(t_0, \tilde{t}_i) \tilde{\delta}_i F_{j}(t_0, \tilde{t}_{i})
\end{equation}
where $P^{D}(t_0, t)$ is a risk-free discount factor at time $t_0$ for maturity $t$ and  $F_{\Delta}(t_0,t_{k})$ is the forward Libor or Euribor rate defined as the fixed rate to be exchanged at time $\tilde{t}_i$ against the $j$-tenor Libor or Euribor rate established at time $\tilde{t}_{i-1}$ so that the swap has zero value at time $t_0$. As in the previous example, the left hand side of \ref{eq:IRS} represents the fixed leg present value whereas the right hand side corresponds to the floating leg present value. For more details, see, for instance \citet{Chibane2009}.

\end{ex}

\begin{ex}[Credit curve based on CDS] 
\label{ex:credit_curve}

Let $S$ be the fair spread of a credit default swap with maturity $T$ and with premium payment dates $t_1<\cdots<t_n=T$. If we denote by $R$ the  expected recovery rate of the reference entity  and by 
$\delta_k$ the year fraction corresponding to period $(t_{k-1}, t_k)$, then 
 the swap equilibrium relation can be expressed as

 
\begin{equation}
\label{eq:CDS}
  S \sum_{k=1}^n \delta_k P^D(t_0,t_k) Q(t_0, t_k)  = -(1-R)\int_{t_0}^T P^D(t_0, t) dQ(t_0,t)
\end{equation}
where $P^{D}(t_0, t)$ is a risk-free discount factor at time $t_0$ for maturity date $t$ and where $t \rightarrow Q(t_0,t)$ is the survival distribution function of the underlying reference entity at time $t_0$. The left hand side of \ref{eq:CDS} represents the premium leg present value whereas the right hand side corresponds to the protection leg present value.
We implicitly assume here that recovery, default and interest rates are stochastically independent.
Using an integration by parts, it is straightfoward to show that survival probabilities $Q(t_0,t)$, $t_0 \leq t\leq T$,  are linked by a linear relation:


\begin{equation}
\begin{split}
S &\sum_{k=1}^n \delta_k P^D(t_0,t_k) Q(t_0, t_k) + (1-R) P^D(t_0,T)Q(t_0,T)\\ 
& \hspace{0.5 cm} + (1-R)\int_{t_0}^T f^D(t_0,t)P^D(t_0, t)Q(t_0,t)dt = 1 - R
\end{split}
\end{equation}
where $f^D(t_0,t)$ is the instantaneous forward rate\footnote{Instantaneous forward rates can be derived from discount factors through the following relation: $f^D(t_0, t)P^D(t_0, t) = - \frac{\partial P}{\partial t} (t_0, t)$.} at time $t_0$ for maturity time $t$.

\end{ex}

\begin{pro}
\label{Pro:convex}
Under Assumption \ref{Ass:linear}, the  set of admissible curves is convex.
\end{pro}

\proof
Convex combination of any admissible curves is an admissible curve.
Indeed, the no-arbitrage requirement (which is a monotonicity condition) and the smoothness condition are preserved by convex combination. The market fit condition is equivalent to impose that some points of the curve are related through a rectangular system of linear equations. If two curves satisfy this linear system, then every convex combination of these two curves also does.
\finproof

The convex nature of the set of admissible curves could be very helpful in measuring this set. It means that if one is able to identify two specific admissible curves, then all possible convex combinations of these two curves are immediately  identified to be admissible. In other word, identifying the set of admissible yield-curves amounts to identify its convex hull, which, under certain conditions, could  be characterized by its extreme points.

\section{Arbitrage-free bounds}
\label{sec:bounds}

As pointed out previously, a term structure building process bears on a series of 
market quotes which are usually reliable only for a small set of standard maturities. For instance, the interest-rate swap market is quite liquid for annual maturities up to 10y and becomes less liquid for higher maturities. 
As for the credit market, CDS contracts are typically considered to be liquid for protection maturities of 3, 5, 7 and 10 years. 
Before addressing the question of curve uncertainty and its impact on pricing and risk management, the first step is to identify the range of values that can be attained by admissible curves. In this section, we restrain ourselves to curve construction processes which both respect  the no-arbitrage and the market-fit constraints. No particular smoothness condition is imposed at this stage in the curve building process. We show how to construct  bounds for the underlying primary quantities at the most liquid maturities. 

\subsection{Bounds for OIS discount factors}


Let us assume that OIS par rates $S_1,\cdots,S_n$ are observed at time $t_0$ for 
standard maturities $T_1<\cdots <T_n$. 
Recall that, for maturities greater than one year, an OIS contract has annual fixed and floating interest payments. We denote by $t_1<\cdots<t_{p_1}<\cdots<t_{p_n}= T_n$ the annual time grid up to the last maturity $T_n$ where $t_0<t_1$ and the index $p_i$ is defined  such that $t_{p_i}=T_i$ for $i=1,...,n$. In other words,  $p_i$ is the index in the payment time grid associated with standard maturity $T_i$. 
We assume that par rates of OIS contracts are reliable for the following maturities: 1 to 10 years, then 15, 20, 25 and 30 years. Note that the payment and the maturity time grids coincide for the first 10 annual maturities, i.e., $t_1=T_1,\cdots, t_{10}=T_{10}$ ($p_i = i$, for $i=1, \ldots,10$). Let $i_{0}$ be the smallest index such that $T_{i_0} \ne t_{i_0}$ ($i_0 = 11$ in our applications). Recall from Example \ref{ex:discount_curve} that, under the market fit condition, the discount factors $P^D(t_0, t_k)$, $k=1, \ldots, p_n$ are connected through the following (rectangular) system of linear equations:
\begin{equation}
\label{eq:OIS_market_fit}
S_i \sum_{k=1}^{p_i-1}  \delta_k P^{D}(t_0, t_k)  + \left(S_i \delta_{p_i} + 1\right) P^{D}(t_0, T_i) = 1\,, \;\;i=1, \ldots, n.
\end{equation}
For $i<i_0$, there is as many equations than unknown discount factors. Then, finding the unknown discount factors amounts to solve a triangular linear system. For $i\ge i_0$, there are less equations than unknown discount factors and a whole set of discount factors can be reached by arbitrage-free curves. 
The next proposition gives the range of values that a market-compatible arbitrage-free curve may attain at standard maturities.

 \begin{pro}
 \label{pro:OIS_bounds}
Assume that, at time $t_0$, quoted OIS par rates $S_1, \ldots, S_n$ are reliable for standard maturities $T_1< \ldots< T_n$.  Recall that $i_0$ corresponds to the index of the first interest-rate payment date which differs from a standard maturity date.
At maturity dates $T_1, \ldots, T_n$, the discount factors associated with market-compatible and arbitrage-free curves are such that:\\
\begin{align}
\label{eq:discount_1}
P^{D}(t_0,T_1) &= \frac{1}{1+S_1\delta_1},\\
\label{eq:discount_2}
P^{D}(t_0,T_i) &= \frac{1}{1+S_i\delta_i}\left(1-\frac{S_i}{S_{i-1}}\left(1-P^D(t_0, T_{i-1})\right)\right),\;\;  i=2, \ldots, {i_0}-1
\end{align}
and, for any $i=i_0, \ldots, n,$
\begin{equation}
\label{eq:bounds_OIS}
P_{\min}^D(t_0, T_i) \le P^{D}(t_0,T_{i}) \le P_{\max}^D(t_0, T_i)
\end{equation}
where
\begin{align}
\label{min_discount}
P_{\min}^D(t_0, T_i) &= \frac{1}{1+S_{i}\delta_{p_{i}}}\left(1-\frac{S_{i}}{S_{i-1}} \left(1-(1-S_{i-1}H_i)P^D(t_0, T_{i-1})\right)\right),\\
\label{max_discount}
P_{\max}^D(t_0, T_i) &= \frac{1}{1+S_{i} (H_i+\delta_{p_i})}\left(1-\frac{S_{i}}{S_{i-1}} \left( 1-P^D(t_0, T_{i-1})\right)\right),
\end{align}
with $H_{i}:=\displaystyle \sum_{k=p_{i-1}+1}^{p_{i}-1}\delta_k$.
 \end{pro}

 \proof 
 For any $i=2, \ldots, n$, line $i-1$ and line $i$ of the linear system (\ref{eq:OIS_market_fit}) can be expressed as
 \begin{align*}
  S_{i-1} \sum_{k=1}^{p_{i-1}}  \delta_k P^{D}(t_0, t_k) &+ P^{D}(t_0, T_{i-1}) = 1,\\
  S_i \sum_{k=1}^{p_{i-1}}  \delta_k P^{D}(t_0, t_k) &+ S_i \sum_{k=p_{i-1}+1}^{p_{i}}  \delta_k P^{D}(t_0, t_k) +  P^{D}(t_0, T_i) = 1.
\end{align*}
 By combining the two previous lines, the $i$-th line of the linear system  (\ref{eq:OIS_market_fit}) can be reformulated as
 \begin{equation}
 \label{eq:bidiag_system}
   \frac{S_i}{S_{i-1}} \left(1- P^D(t_0, T_{i-1})\right) + S_i \sum_{k=p_{i-1}+1}^{p_{i}-1}  \delta_k P^{D}(t_0, t_k) +   \left(1+ S_i\delta_{p_i}\right)P^{D}(t_0, T_i) = 1.
 \end{equation}
We then remark that, for $i=1, \ldots, i_0-1$, the linear system (\ref{eq:OIS_market_fit}) is bidiagonal and can  trivially be solved recursively. This boils down to equations (\ref{eq:discount_1}) and (\ref{eq:discount_2}). 
The bounds (\ref{eq:bounds_OIS}) are obtained by considering that arbitrage-free discounting curves are nonincreasing, so that  between any two successive standard maturities $T_{i-1}$ and $T_i$, $i=i_0, \ldots, n,$ discount factors must be greater than the value taken at $T_i$ and smaller than the value taken at $T_{i-1}$.
Consequently, the minimum and maximum discount values $P_{\min}^D(t_0, T_i)$ and $P_{\max}^D(t_0, T_i)$ given respectively by (\ref{min_discount}) and (\ref{max_discount}) can immediately be derived from equation (\ref{eq:bidiag_system}) and the fact that, in absence of arbitrage opportunity, 
$$
P^{D}(t_0,T_{i}) \le P^{D}(t_0,t_{k}) \le P^{D}(t_0,T_{i-1}),  
$$ 
for $k = p_{i-1}+1, \ldots, p_i - 1$.
\finproof

\begin{rem}
Note that, as soon as $i>i_0$, these bounds cannot be computed explicitly since $P_{\min}^D(t_0, T_i)$ and $P_{\max}^D(t_0, T_i)$ may depend on the unknown discount factors $P^D(t_0, T_k)$, $k=i_0, \ldots, i$. However, these bounds can be used together with a bootstrap procedure to successively limit  the exploration set of possible discount factors when stripping the curve at each standard maturities.
\end{rem}
We now give a recursive algorithm which allows to obtain  bounds that do not depend on any building process (model-free bounds).
For a particular maturity time $T_i$, the unknown discount factors in expressions (\ref{min_discount}) and  (\ref{max_discount}) can be replaced by the worst bounds computed at the preceding date $T_{i-1}$.

 \begin{pro}
 \label{pro:OIS_bounds_model_free_sharp}
Assume that, for any $i=i_0, \ldots,n$, the quantity $1-S_{i-1}H_i$ is positive.
The following recursive procedure provides model-free bounds for OIS discount factors at standard maturities. 
\begin{itemize}
\item \textbf{Step 1}: For $i=1, \ldots, {i_0}-1$, compute recursively 
$P^{D}(t_0,T_i)$ using equations (\ref{eq:discount_1}) and (\ref{eq:discount_2}). 
\item \textbf{Step 2}: For $i=i_0, \ldots, n$,
\begin{equation}
\label{eq:MF_bounds_OIS}
P_{\min}(T_i) \le P^{D}(t_0,T_{i}) \le P_{\max}(T_i)
\end{equation}
where
\begin{align}
\label{MF_min_discount}
P_{\min}(T_i) &= \frac{1}{1+S_{i}\delta_{p_{i}}}\left(1-\frac{S_{i}}{S_{i-1}} \left(1-(1-S_{i-1}H_i)P_{\min}(T_{i-1}) \right)\right),\\
\label{MF_max_discount}
P_{\max}(T_i) &= \frac{1}{1+S_{i} (H_i+\delta_{p_i})}\left(1-\frac{S_{i}}{S_{i-1}} \left( 1-P_{\max}(T_{i-1})\right)\right).
\end{align}
\end{itemize}
In addition, the bounds are sharp in the sense that the set of lower bounds $(P_{\min}(T_i))_i$ is reached by a market-compatible arbitrage-free curve $P_{\min}(t_0, \cdot)$ and the set of upper bounds $(P_{\max}(T_i))_i$ is reached by another market-compatible arbitrage-free curve $P_{\max}(t_0, \cdot)$. 
 \end{pro}

\proof
If, for any $i=i_0, \ldots, n$, the quantity $1-S_{i-1}H_i$ is positive, the left hand sides of expressions (\ref{min_discount}) and (\ref{max_discount}) correspond to an increasing function of $P^D(t_0, T_{i-1})$. The minimum (resp. maximum) value is attained at $T_i$ for $P^D(t_0, T_{i-1}) = P_{\min}(T_{i-1})$  (resp. for $P^D(t_0, T_{i-1}) = P_{\max}(T_{i-1})$). The bounds (\ref{MF_min_discount}) and (\ref{MF_max_discount}) are sharp since they are reached by some ``extreme'' arbitrage-free curves. More specifically, the values $(P_{\min}(T_i))$ are reached at times $T_i$, $i=i_0, \ldots, n$ by a curve $P_{\min}(t_0, \cdot)$ such that, for any $i=1, \ldots, i_0-1$, $P_{\min}(t_0, T_i)$ is defined by (\ref{eq:discount_1}-\ref{eq:discount_2}) and  for any $i=i_0, \ldots, n$, $P_{\min}(t_0, t) = P_{\min}(T_{i-1})$, $T_{i-1} \leq t < T_i$. The values $(P_{\max}(T_i))$ are reached at times $T_i$, $i=i_0, \ldots, n$ by a curve $P_{\max}(t_0, \cdot)$ such that, for any $i=1, \ldots, i_0-1$, $P_{\max}(t_0, T_i)$ is defined by (\ref{eq:discount_1}-\ref{eq:discount_2}) and  for any $i=i_0, \ldots, n$, $P_{\max}(t_0, t) = P_{\max}(T_{i})$, $T_{i-1} < t \leq T_i$.
\finproof

\begin{rem}
To derive equation \ref{MF_min_discount}, we implicitly assume that, for any $i=i_0, \ldots, n$, the quantity $1-S_{i-1}H_i$ is positive, so that the left hand side of expression (\ref{min_discount}) corresponds to an increasing function of $P^D(t_0, T_{i-1})$. For the OIS-based construction of discounting curve, $H_i$ is typically smaller than $10$ (no more than 10 years between two consecutive standard maturities) and this assumption is satisfied for all standard maturities as soon as the OIS par rates $S_i$ are smaller than $10\%$.
\end{rem}

\begin{rem}
Note that, for any $i=i_0, \ldots, n$, every value in the interval $(P_{\min}(T_i), P_{\max}(T_i))$ is reached by a particular arbitrage-free curve which fits market quotes. This is because a particular value $P_i$ in  $(P_{\min}(T_i), P_{\max}(T_i))$ is characterized by a convex combination of $P_{\min}(T_i)$ and $P_{\max}(T_i)$, i.e.,  there exists $\alpha$ in $(0,1)$ such that $P_i = \alpha P_{\min}(T_i) + (1-\alpha)P_{\max}(T_i)$. The curve $P^D_{\alpha}$ defined by $\alpha P_{\min}(t_0, \cdot) +  (1-\alpha)P_{\max}(t_0, \cdot)$ goes through the point $(T_i, P_i)$ and, in addition, it fits market quotes and is arbitrage-free since the two latter properties are invariant by convex combinations. 
\end{rem}

Figure \ref{fig:OIS_bounds} displays, for each standard maturity, the no-arbitrage set of discount factors (top graph) and the corresponding set of continuously-compounded spot rate curves (bottom graph) that are perfectly compatible with quoted OIS par rates as of May 31st, 2013 (given in Table \ref{table:OIS_data}). Recall that OIS discount factors can be computed without uncertainty from maturity spanning from 1y to 10y. For the next maturities, sharp model-free bounds can be computed using the algorithm described in Proposition \ref{pro:OIS_bounds_model_free_sharp}. The bounds associated with time-to-maturities 15y, 20y, 30y and 40y are represented by a black segment. We also plot the curves $P_{\min}(t_0, \cdot)$ (in solid line) and $P_{\max}(t_0, \cdot)$ (in dashed line) which match (resp.) the lower and the upper bounds at these maturities. Note that between two standard maturities, the range in discount rates can be greater than one point of percentage.\\

\begin{figure}[h]
\begin{center}
\includegraphics[width = 0.7\linewidth, height = 5.5cm]{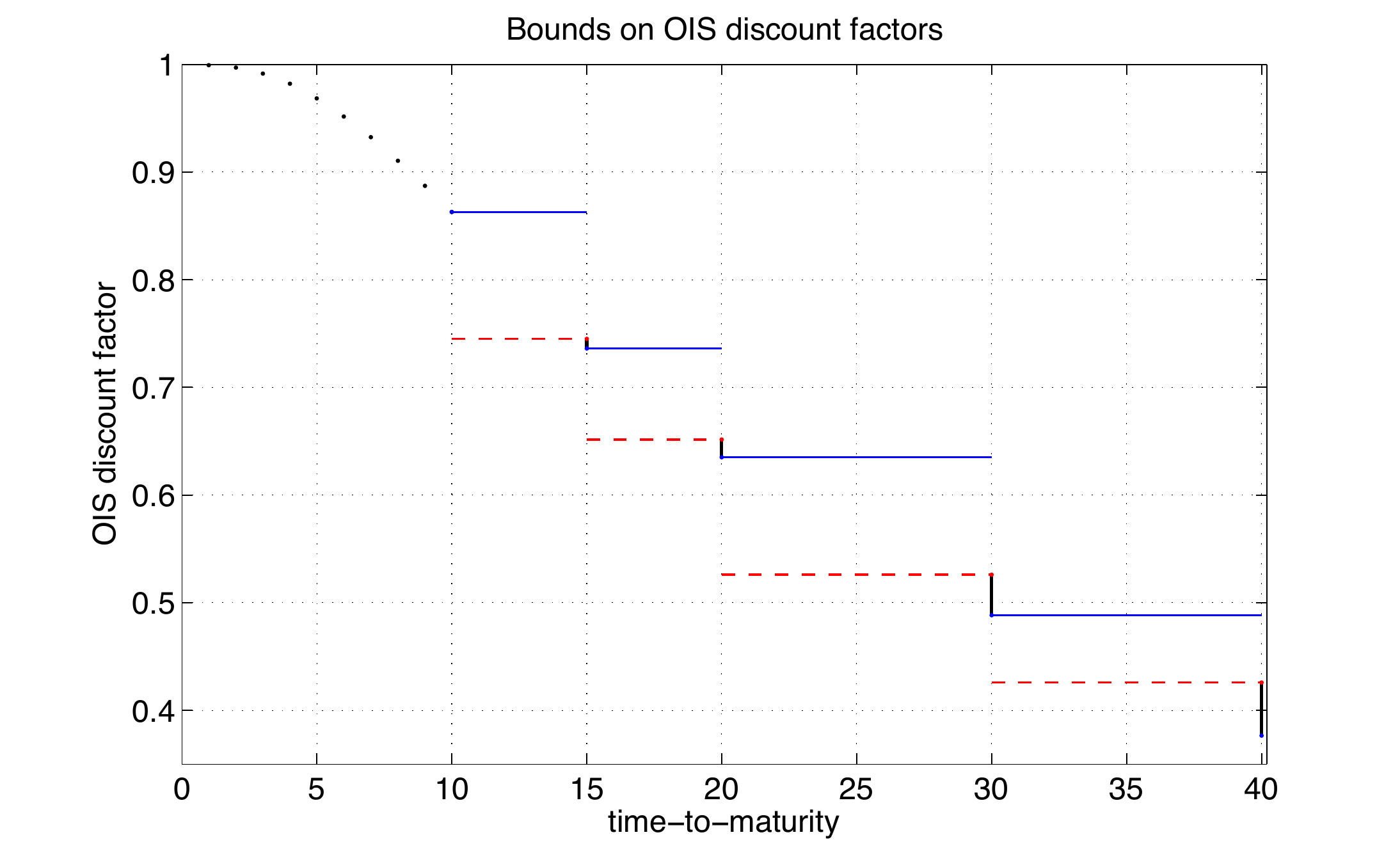}
\includegraphics[width = 0.7\linewidth, height = 5.5cm]{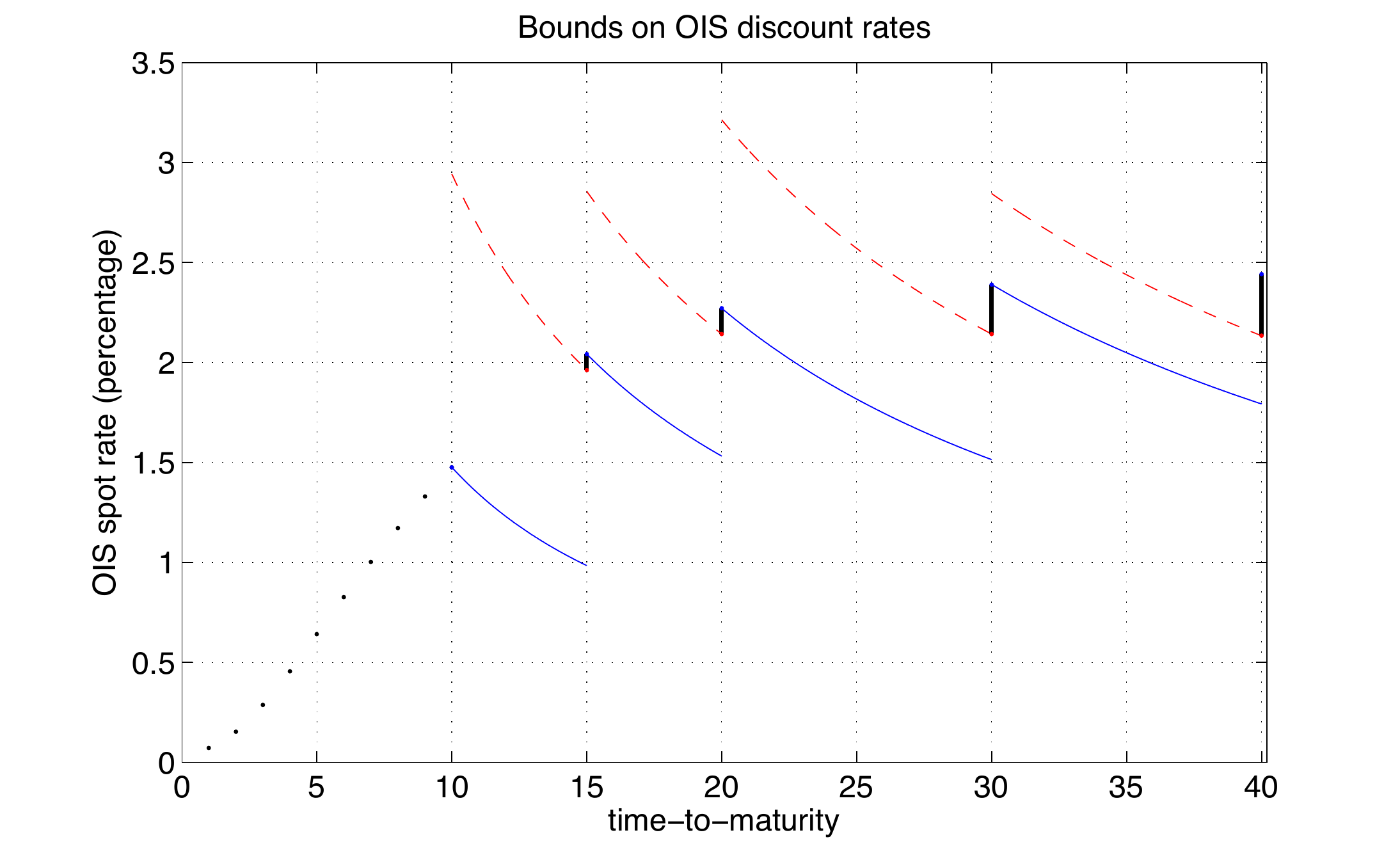}
\caption{Upper graph: bounds on discount factors constructed  from OIS par rate as of May 3st, 2013. The solid lines correspond to the curve   $P_{\min}(t_0, \cdot)$ which reaches the lower bounds (\ref{MF_min_discount}) whereas dashed lines correspond to the curve $P_{\max}(t_0, \cdot)$ which reaches the upper bounds (\ref{MF_max_discount}).
Lower graph: corresponding bounds and ``extreme'' curves for continuously-compounded spot rates.}
\label{fig:OIS_bounds}
 \end{center}
\end{figure}

\begin{table}[h]
\centering
\begin{tabular}{|c|c|c|c|c|c|c|c|}
\hline 
maturity (year) & $1$ & $2$ & $3$ & $4$ & $5$ & $6$ & $7$\\
\hline
swap rate (percentage) & $0.0720$ & $0.1530$ & $0.2870$ & $0.4540$ & $0.6390$ & $0.8210$ & $0.9930$\\
\hline
\hline
maturity (year) & $8$ & $9$ & $10$ & $15$ & $20$ & $30$ & $40$\\
\hline
swap rate (percentage) & $1.1570$ & $1.3090$ & $1.4470$ & $1.9300$ & $2.1160$ & $2.1820$ & $2.2090$\\
\hline
\end{tabular}
\caption{OIS swap rates as of May 31st, 2013}
\label{table:OIS_data}
\end{table}

Another application of  Proposition \ref{pro:OIS_bounds_model_free_sharp}
 is to identify  a union of rectangles $\bigcup_{i=1}^{n} \mathcal{R}_i
$ in which any arbitrage-free and perfect-fit discounting curve must lie. Given that arbitrage-free discount factors are nonincreasing, these rectangles are defined by the following couple of (bottom-left, top-right) points: $\mathcal{R}_1 = \{(0,P^D(t_0, T_1)), (T_1, 1)\}$, $\mathcal{R}_i = \{(T_{i-1},  P^D(t_0, T_{i})),$ $(T_{i},  P^D(t_0, T_{i-1})\}$ for  $i=2, \ldots, i_0-1$ and $\mathcal{R}_i$ $=$ $\{(T_{i-1},$ $P_{\min}(T_{i})),$ $(T_{i},$  $P_{\max}(T_{i-1})\}$, for $i=i_0, \ldots, n$. For  OIS quoted par rates as of May 31st, 2013, the union of rectangles is displayed in Figure \ref{fig:OIS_tunnel}.\\ 

\begin{figure}[h]
\begin{center}
\includegraphics[width = 0.7\linewidth, height = 6cm]{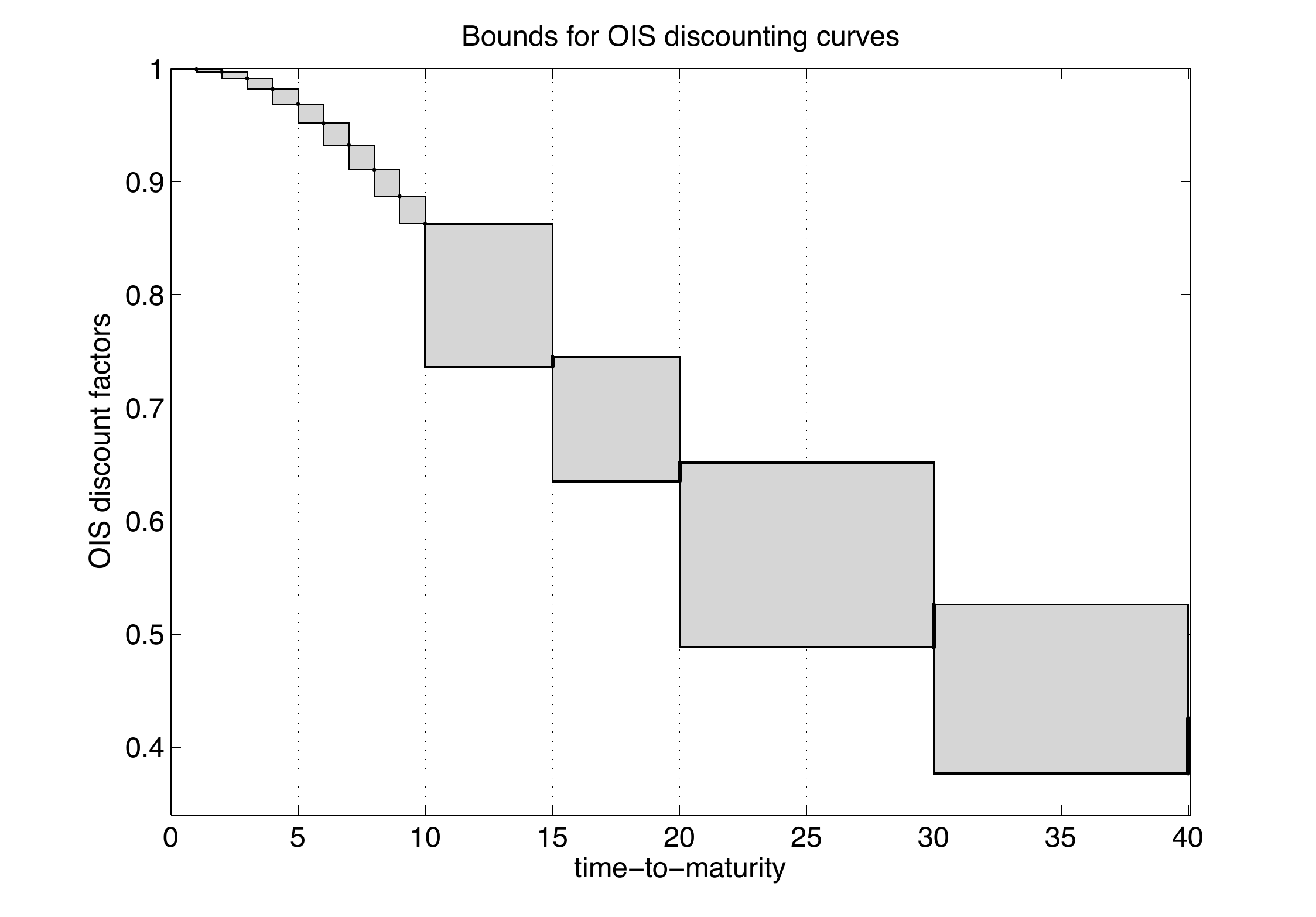}
\caption{Union of rectangles in which any arbitrage-free discounting curve must lie when fitted on quotes OIS par rate as of May 31st, 2013.}
\label{fig:OIS_tunnel}
 \end{center}
\end{figure}

%
%

Note that  Proposition \ref{pro:OIS_bounds_model_free_sharp} can also be used to detect whether arbitrage opportunities are hidden in market data. The set of market quotes $(S_i)_{i=1, \ldots, n}$ is arbitrage-free if, for any $i=1, \ldots, i_0-1$,  $P^{D}(t_0,T_i)$ are such that $P^{D}(t_0,T_i) < P^{D}(t_0,T_{i-1})$ where $P^{D}(t_0,T_{0}) := 1$ and if for any $i=i_0, \ldots, n$, $P^{D}_{\min}(t_0,T_i) < P^{D}_{\max}(t_0,T_{i})$. The following proposition gives a method to detect arbitrage opportunities in the set of quoted OIS par rates.

 \begin{pro}
 \label{pro:OIS_no_arbitrage}
 Assume that OIS discount factors and model-free bounds are computed by launching the recursive algorithm described in Proposition \ref{pro:OIS_bounds_model_free_sharp}. An arbitrage opportunity can be detected in the data set $(S_i)_{i=1, \ldots, n}$ at the first index $i$ such that   
\begin{equation}
\label{eq:cond_1}
S_i < \left( \frac{1}{S_{i-1}} + \delta_{i}\frac{P^D(t_0, T_{i-1})}{1-P^D(t_0, T_{i-1})} \right)^{-1}, \; i=2, \ldots, i_0-1,
\end{equation}
\begin{equation}
\label{eq:cond_2}
S_i < \left( \frac{1}{S_{i-1}} + (H_i + \delta_{p_i})\frac{P_{\max}(T_{i-1})}{1-P_{\max}(T_{i-1})} \right)^{-1}, \; i=i_0, \ldots, n.
\end{equation}
 \end{pro}
 
 \proof
 If the first index $i$ is between $2$ and $i_0-1$, thanks to (\ref{eq:discount_2}) inequality (\ref{eq:cond_1}) leads to $P^{D}(t_0,T_i) > P^{D}(t_0,T_{i-1})$. If the first index $i$ is bewteen $i_0$ and $n$,  inequality (\ref{eq:cond_2}) yields $P^{D}_{\min}(t_0,T_i) > P^{D}_{\max}(t_0,T_{i})$ and, in that case, there is no nonincreasing curves on $(T_{i-1}, T_i)$ that fits market rate $S_i$ at time $T_i$. \finproof

\begin{rem}
As a consequence of Proposition \ref{pro:OIS_no_arbitrage}, an increasing sequence of OIS par rates $S_1\leq \cdots \leq S_n$ is always associated with an arbitrage-free curve.
\end{rem}


\subsection{Bounds for risk-neutral survival probabilities}

In the previous subsection, we present a methodology for building  market-consistent arbitrage-free bounds for OIS discount factors at standard maturities. Here, the same approach is adapted to the case of credit curves, i.e., term-structure of default probabilities implied from a set of CDS spreads.\\

We consider a particular underlying entity (corporate or sovereign issuer) or a credit index on which CDS protection is quoted at several maturities. Let us assume that, for this entity, CDS fair spreads $S_1,\cdots,S_n$  are observed at  time $t_0$ for standard protection maturities $T_1<\cdots <T_n$.  We denote by $t_1<\cdots<t_{p_1}<\cdots<t_{p_n}= T_n$ the premium payment dates where  $t_0<t_1$ and the set of indices $(p_i)$ is such that $t_{p_i}=T_i$ for $i=1,...,n$.\\ 

 Recall from Example \ref{ex:credit_curve} that, under the market fit condition, the survival probabilities $Q(t_0, t_k)$, $k=1, \ldots, p_n$ are connected through the following system of linear equations
\begin{equation}
\label{eq:credit_system}
\begin{split}
S_i &\sum_{k=1}^{p_i} \delta_k P^D(t_0,t_k) Q(t_0, t_k) + (1-R) P^D(t_0,T)Q(t_0,T)\\ 
& \hspace{0.5 cm} + (1-R)\int_{t_0}^{T_i} f^D(t_0,t)P^D(t_0, t)Q(t_0,t)dt = 1 - R\,, \;\;i=1, \ldots, n.
\end{split}
\end{equation}
where $R$ is the expected recovery rate and $f^D(t_0,t)$ is the instantaneous forward rate associated with the discounting curve $P^D(t_0, \cdot)$. The next proposition gives the range of survival probabilities that a market-compatible arbitrage-free credit curve may attain at standard maturities.

\begin{pro}
Assume that, at time $t_0$, quoted fair spreads $S_1, \ldots, S_n$ are reliable for standard CDS maturities $T_1< \ldots< T_n$.  For any $i=1, \ldots,n$, the survival probability $Q(t_0,T_{i})$ associated with a market-compatible and arbitrage-free credit curve is such that:\\
\begin{equation}
\label{bounds_survi}
Q_{\min}(t_0, T_i) \le Q(t_0,T_{i}) \le Q_{\max}(t_0, T_i)
\end{equation}
where
\begin{align}
\label{min_survi}
Q_{\min}(t_0, T_i) = \frac{1-R -  \displaystyle \sum_{k=1}^{i} \left( (1-R)M_k + S_i N_k \right)Q(t_0,T_{k-1}) }{P^D(t_0,T_i)(1-R+S_i\delta_{p_i})},\\
\label{max_survi}
Q_{\max}(t_0, T_i) = \frac{1-R -  \displaystyle \sum_{k=1}^{i-1} \left( (1-R)M_k + S_i N_k \right)Q(t_0,T_{k}) }{P^D(t_0,T_{i-1})(1-R) + S_i\left(N_i + \delta_{p_i}P^D(t_0,T_{i})\right)},
\end{align}
with $p_0:=1$, $T_0:=t_0$ (thus $P^{D}(t_0,T_{0}) = Q(t_0,T_{0}) = 1$) and, for any  $i=1, \ldots,n$, $M_i := P^{D}(t_0,T_{i-1})-P^{D}(t_0,T_{i})$  and $N_i:=\displaystyle \sum_{k=p_{i-1}}^{p_i-1}  \delta_k P^{D}(t_0,t_k)$.
\end{pro}
\proof The bounds (\ref{bounds_survi}) are obtained by considering that an arbitrage-free term-structure of survival probabilities must be nonincreasing, so that  between any two successive standard maturities $T_{i-1}$ and $T_i$, $i=i_0, \ldots, n,$ survival probabilities should be greater than the value taken at $T_i$ and smaller than the value taken at $T_{i-1}$.
Consequently, the minimum and maximum survival probability values $Q_{\min}(t_0, T_i)$ and $Q_{\max}(t_0, T_i)$ given respectively by (\ref{min_survi}) and (\ref{max_survi}) can immediately be derived from equation (\ref{eq:credit_system}) and the following  system of inequalities :
\begin{equation}
\left\lbrace 
\begin{array}{lll} 
  Q(t_0,T_{1}) \le Q(t_0, t) \le 1 & \mbox{for} \ t_0 \le t < T_1,\ \\ 
\hspace{2cm} \vdots & \\
  Q(t_0,T_{i}) \le Q(t_0, t) \le Q(t_0,T_{i-1}) & \mbox{for} \ T_{i-1} \le t < T_i
\end{array}\right.
\end{equation}
\finproof

\begin{rem}
Note that these bounds cannot be computed explicitly since, for every $i=1,\ldots,n$, the lower and upper bounds (\ref{min_survi}) and (\ref{max_survi}) depend on the survival probabilities $Q(t_0, T_k)$, $k=1, \ldots, i-1$ and the latter probabilities are not known with certainty. However, as for the OIS-based discounting curves, these bounds can be used together with a bootstrap procedure to successively limit  the exploration set of possible survival probabilities when stripping the curve at each standard maturities.
\end{rem}

For a particular maturity time $T_i$, the unknown survival probabilities $Q(t_0, T_k)$, $k=1, \ldots, i-1$  in expressions  (\ref{min_survi}) and (\ref{max_survi}) can be replaced by the worst bounds computed at the preceding steps  $k=1, \ldots, i-1$. 
This argument leads to the construction of an iterative procedure that allows to compute model-free bounds for survival probabilities at each standard CDS maturities. This procedure is given in the following proposition.

 \begin{pro}
 \label{pro:CDS_survi_model_free}
For each standard CDS maturity, model-free bounds for implied survival probabilities can be computed using the following recursive procedure.\\ 

\noindent For $i=1, \ldots, n$, compute recursively 
\begin{equation}
\label{eq:MF_bounds_CDS}
Q_{\min}(T_{i}) \le Q(t_0,T_{i}) \le Q_{\max}(T_{i})
\end{equation}
where
\begin{align}
\label{MF_min_survi}
Q_{\min}(T_{i}) & = \frac{1-R -  \displaystyle \sum_{k=1}^{i} \left( (1-R)M_k + S_i N_k \right)Q_{\max}(T_{k-1}) }{P^D(t_0,T_i)(1-R+S_i\delta_{p_i})}
\\
\label{MF_max_survi}
Q_{\max}(T_{i}) & = \frac{1-R -  \displaystyle \sum_{k=1}^{i-1} \left( (1-R)M_k + S_i N_k \right)Q_{\min}(T_{k}) }{P^D(t_0,T_{i-1})(1-R) + S_i\left(N_i + \delta_{p_i}P^D(t_0,T_{i})\right)}
\end{align}
and $Q_{\max}(T_{0}):=1$.
 \end{pro}

As for the OIS-based discount curve construction,  Proposition  \ref{pro:CDS_survi_model_free} can be used to identify  a union of rectangles $\bigcup_{i=1}^{n} \mathcal{R}_i
$ in which any arbitrage-free and perfect-fit credit curve must lie. Indeed, given that any arbitrage-free term-structure of survival probabilities is nonincreasing, these rectangles are defined by the following couple of (bottom-left, top-right) points: $\mathcal{R}_i$ $=$ $\{(T_{i-1},$ $Q_{\min}(T_{i})),$ $(T_{i},$  $Q_{\max}(T_{i-1})\}$, for $i=1, \ldots, n$ where $T_0 = t_0$ and $Q_{\max}(T_{i-1}) = 1$. For   CDS spreads of AIG as of Dec. 17, 2007 (given in Table \ref{CDS_data}), the union of rectangles is displayed in Figure \ref{fig:CDS_tunnel} where for all standard maturities, the model-free bounds (\ref{eq:MF_bounds_CDS}) have been represented by black segments.

\begin{figure}[h]
\begin{center}
\includegraphics[width = 0.7\linewidth, height = 6cm]{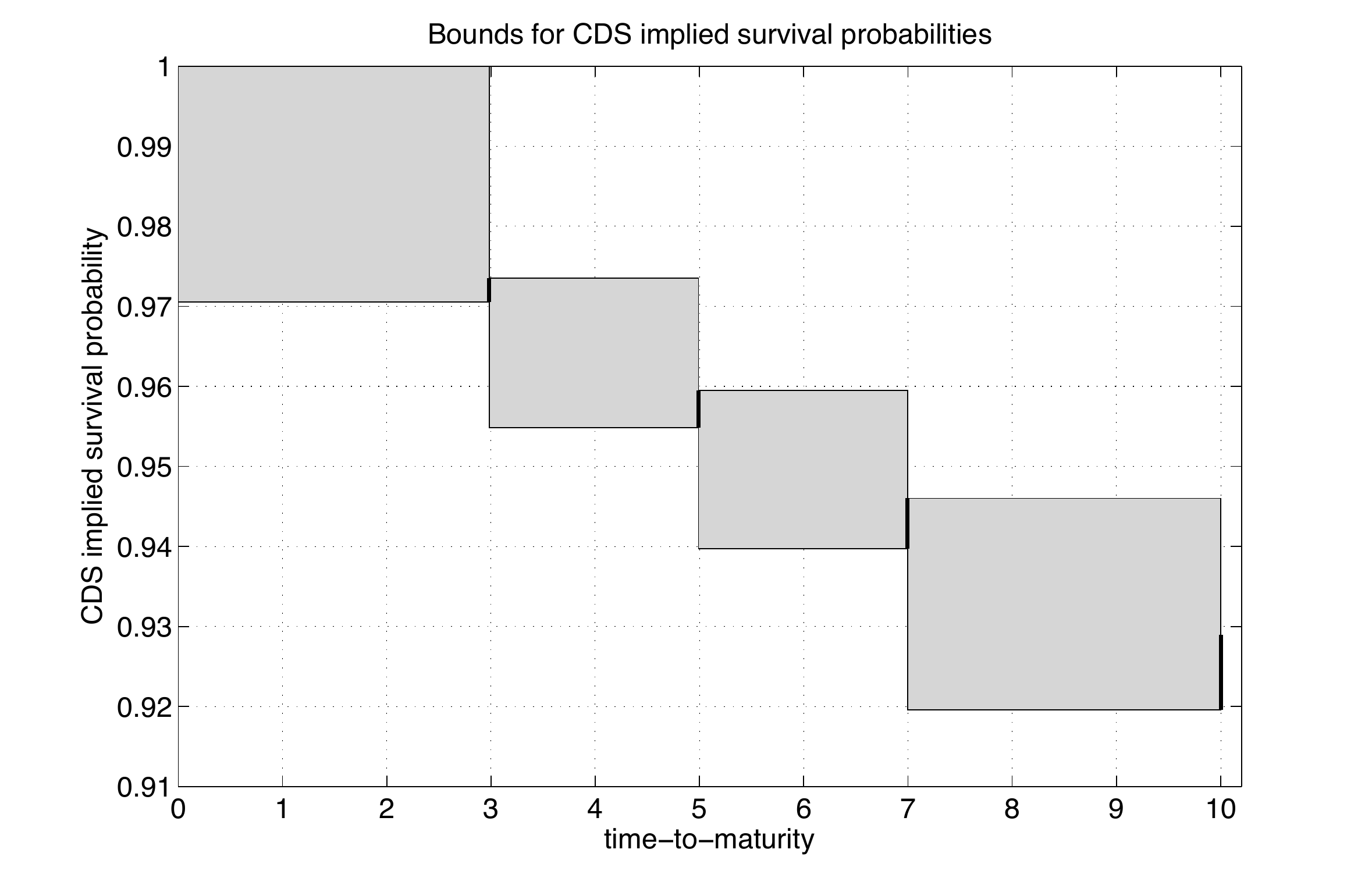}
\caption{Union of rectangles in which any arbitrage-free survival curve must lie when fitted on CDS spreads of AIG as of Dec 17, 2007. Survival probability  bounds (\ref{MF_min_survi}) and (\ref{MF_max_survi}) have been computed with $R=40\%$ and a discounting curve such that $P^D(t_0, t) = \exp(-3\% (t-t_0))$.}
\label{fig:CDS_tunnel}
 \end{center}
\end{figure}

\begin{table}[h]
\centering
\begin{tabular}{|c|c|c|c|c|}
\hline 
maturity (year) & $3$ & $5$ & $7$ & $10$ \\
\hline
CDS spread (bp) & $58$ & $54$ & $52$ & $49$ \\
\hline
\end{tabular}
\caption{AIG CDS spread at Dec. 17, 2007}
\label{CDS_data}
\end{table}


\newpage

 Figure \ref{fig:Bounds_evol} shows the sensitivity of the lower and upper bounds with respect to the recovery rate assumption. As expected, for any standard maturities, the bounds are decreasing function of the recovery rate. This is consistent with the fact that, when the expected loss in case of default decreases, the default probability has to increase in order to reach the same level of CDS spread. Interestingly, the size of the bounds, which can be interpreted as a measure of uncertainty, increases with the recovery rate assumption. For expected recovery lower than $40\%$, the recovery impact on the size of the bounds can be considered to be insignificant whereas this is no longer the case  for recovery larger than $40\%$.
 
 \begin{figure}[h]
\begin{center}
\includegraphics[height=5.3cm, width = 12cm]{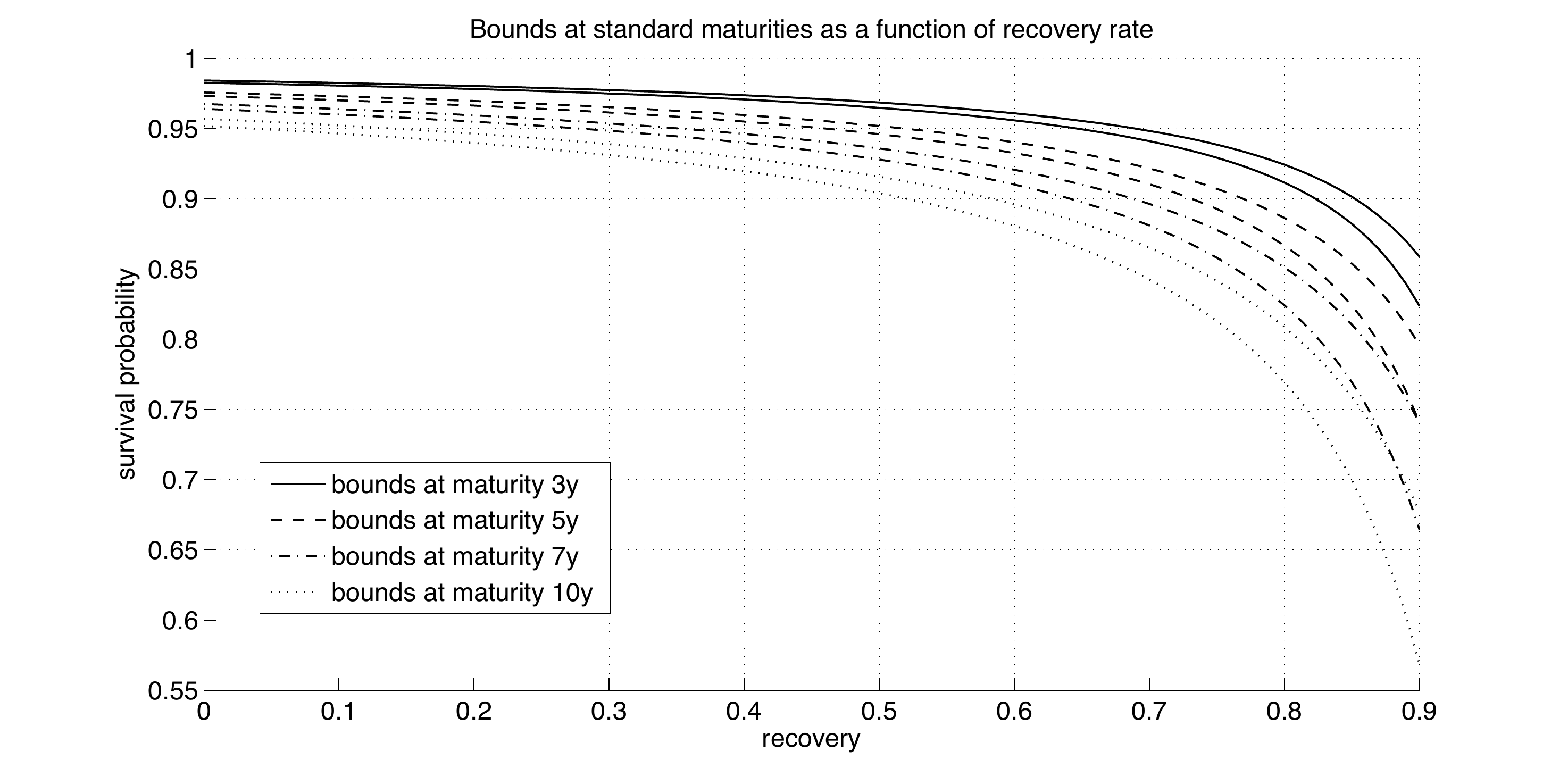}
\caption{Survival probability bounds at standard maturities computed from CDS spreads of AIG as of Dec 17, 2007. The discounting curve is such that $P^D(t_0, t) = \exp(-3\% (t-t_0))$.}
\label{fig:Bounds_evol}
 \end{center}
\end{figure}

\section{Construction of admissible term-structures}
\label{sec:construct_admissible_curve}

In the previous section, we explained how to compute bounds at standard maturities for arbitrage-free and market-consistent discounting or credit curves, given a set of market quotes for some liquidly traded instruments. In particular, we did not impose any smoothness condition on the reconstructed curves, which leads to identify ``extreme curves'' with unrealistic behaviors. In this section, we additionally  force the reconstructed curve to be sufficiently regular and thus to be admissible in the sense of  
Definition \ref{Def:admissible}.
The proposed construction of admissible curves is based on the idea that the class of  dynamic term-structure models is rich enough to generate admissible term-structures.
The constructed term-structures are given as a by-product of a mean-reverting affine model. For instance, an interest-rate curves (swap curves or bond yield curves) will be defined as the initial term-structure of zero-coupon prices  obtained in a short-rate model. In the same vein, a credit curves will be defined as an initial term-structure of survival probabilities obtained in a default intensity model.
The curve building process relies on a piecewise-constant specification of the long-term mean parameter which allows to perfectly reproduce market prices of vanilla interest-rate products (bonds, swaps) and CDS contracts while preserving a certain degree of smoothness.\\


In the sequel, $Y$ will denote a  L{\'e}vy process and $W$ a standard brownian motion. We assume that all introduced processes are defined with respect to a stochastic basis $(\Omega,\mathcal{F},\mathbb{F},\mathbb{Q})$. 
For every considered  L{\'e}vy process $Y$, its cumulant function is denoted by $\kappa$, i.e., $\kappa(\theta)=\log\mathbb{E}\left[e^{\theta Y_1}\right]$ and its set of parameters is denoted by $\mathbf{p}_L$. As a matter of example, some cumulant functions are given in Table \ref{Table:Levy} for the Brownian motion and for two class of L{\'e}vy subordinators parametrized by a single variable $\lambda$ which inversely controls the jump size of the L\'evy process.

		
\begin{table}[h]	
\begin{center}
{\renewcommand{\arraystretch}{1.8}
\begin{tabular}{|l|l|l|}
\cline{2-3} \multicolumn{1}{c|}{}  & L{\'e}vy measure   & Cumulant \\
		\hline
		 Brownian motion &  $\rho(dx)=0 $ &  $\kappa(\theta)=\frac{\theta^2}{2}$  \\

		\hline
		 Gamma process &  $\rho(dx)=\frac{e^{-\lambda x}}{x}1_{x>0}dx$ &  $\kappa(\theta)=-\log\left(1- \frac{\theta}{\lambda}\right)$  \\
		 \hline
		 Inverse Gaussian process&   $\rho(dx)=\frac{1}{\sqrt{2 \pi x^3}}\exp\left(-\frac{1}{2} \lambda^2 x \right)1_{{x>0}}dx$\ & $\kappa(\theta)= \lambda-\sqrt{\lambda^2-2\theta}$   \\
		 \hline
		\end{tabular}
		}
\end{center}
\caption{Examples of L\'evy measures and cumulants}
\label{Table:Levy}	
\end{table}

We refer the reader to \citet{Cont2003} for more details on L\'evy processes. The term-structure curve at time $t_0$ is built from a set of market quotes $\mathbf{S}=(S_1, \cdots, S_n)$ observed at time $t_0$ and associated with an increasing set of maturity dates $\mathbf{T}=(T_1,\cdots,T_n)$. For any $i=1, \ldots,n$, the market quote $S_i$ corresponds to the market price of a financial instrument with maturity date $T_i$.\\

  We assume that, under the risk-neutral probability measure $\Q$ and depending on the type of curve under construction, the  short-term interest rate or the default intensity is either governed by  an extended L\'evy-driven Ornstein-Uhlenbeck process (L\'evy-driven OU)
  \begin{equation}
\label{bra}
dX_t = a(b(t) - X_t)dt + \sigma dY_{ct},
\end{equation}
or an extended CIR process  
\begin{equation}
\label{cir}
dX_t = a(b(t) - X_t)dt + \sigma \sqrt{X_t} dW_{t},
\end{equation}
where the long-term mean parameter $b$ is assumed to be a deterministic function of time, $a$ is a positive parameter which controls the speed of mean-reversion and $\sigma$ is a positive volatility parameter. 
Concerning the L\'evy-driven OU specification \ref{bra}, we use an additional positive parameter $c$ 
which appears as an increasing  change of time $t \rightarrow ct$. This parameter can also be interpreted as a volatility parameter but, contrarily to $\sigma$, it controls jump frequency (an increase of $c$ makes the underlying L\'evy process jumps more frequently). Let $X_0$ be the value at time $t_0$ of the process $X$.  The use of L\'evy processes as a driver of short rate or default intensity dynamics stems from the fact that processes driven by some L\'evy processes provide better fit on time series of bond returns than when driven by a Brownian motion. For more details on the subject of term structure or credit risk modeling with L\'evy processes, the reader is referred for instance to  \citet{Eberlein1999}, \citet{Cariboni2004}, \citet{Kluge2005}, \citet{Crepey2012}.\\

\begin{rem}
Specification (\ref{bra}) corresponds to a L\'evy Hull-White extended Vasicek model but where the focus is put on curve construction instead of curve projection. In the seminal Hull and White approach (see \citet{Hull1990}), the initial term-structure is given as a model input  and the function $b$ is defined in such a way that the input term-structure is reproduced by the model. In our approach, contrary to the Hull and White framework, the deterministic function $b$ is directly calibrated on the set of market quotes.
\end{rem}

The proposed construction of admissible curves is based on a piecewise-constant specification of the long-term mean parameter $b$, i.e,
\begin{equation}
\label{eq:piecewise_const}
 b(t)= b_i,\;\; \mbox{for } T_{i-1} \le t < T_i,\;\;i=1,\ldots,n.
\end{equation}
where $T_0:=t_0$.
This specification is motivated by the following arguments.
\begin{itemize}
\item Interest-rate curves as, for instance, bond yield curves or OIS discount rate curves can be derived from a term-structure of zero-coupon prices. 
Credit curves can be assimilated to a term-structure of survival probabilities. 
As we will see in Subsection \ref{subsec:pricing}, analytical expressions exist for these term-structures when they are computed in models (\ref{bra}) and (\ref{cir}).
\item In addition, the piecewise-constant function (\ref{eq:piecewise_const}) has discontinuity points corresponding to standard maturities $T_1, \ldots, T_n$. We will see in Subsection \ref{subsec:admissible} that this feature allows to transform the over-parameterized  linear system of present values into a triangular system of non-linear equations which can be solved iteratively. The no-arbitrage requirement is guaranteed under some conditions on the implied levels $b_i$, $i=1, \ldots, n$ and the curves obtained after calibration satisfy the smoothness condition  of Definition \ref{def:smooth}.
\end{itemize}


 

\subsection{Curve explicit analytical expressions}
\label{subsec:pricing}

We rely on a standard pricing framework where, in absence of arbitrage opportunity, the value at time $t_0$ of a default-free zero-coupon bond with maturity time $t$ is given by
\begin{equation}
\label{ZC_price}
P(t_0, t) = \E_{\Q}\left[ \exp\left(-\int_{t_{0}}^t{ X_u du}\right) \mid \F_{t_{0}}\right],
\end{equation}
where $\F$ is the natural filtration of the short-rate process $X$. 
Note that (\ref{ZC_price}) is also the expression of the survival probability $Q(t_0, t) = \Q(\tau > t \mid \F_{t_{0}})$ when $X$ is the risk-neutral default intensity of the default time $\tau$.  
Let us denote by $P(t_0, t)$ the value at time $t_0$ of a generic elementary quantity with maturity date $t$. Depending on the type of curve under construction, this quantity can be either the price of a zero-coupon bond in a short-rate model or the survival probability of a particular reference entity in a default intensity model.
When the mean-reverting level $b$ is a deterministic function of time, the following lemma, which is a classical result in the theory of affine term-structure models, gives an analytical expression for $P(t_0, t)$ in the class of L\'evy-driven OU models.

\begin{lem}\label{Levy-OU:affine}
In the L\'evy-driven OU model (\ref{bra}), the value at time $t_0$ of a generic elementary quantity with maturity $t$ is given by
\begin{equation}
\label{eq1}
P(t_0,t) =\exp\left(-\phi(t-t_0)X_0- a\int_{t_0}^{t}{b(u)\phi(t-u)du} -c\psi(t-t_0)
\right)
\end{equation}
where the functions $\phi$ and $\psi$ are defined by
\begin{align}
\label{eq:phi}
\phi(s)&:=\frac{1}{a}\left(1-e^{-as}\right), \\
\label{eq:psi}
\psi(s)&:=-\int_{0}^{s}{\kappa \left(-\sigma\phi(s-\theta)\right)d\theta}.
\end{align}
\end{lem}

\proof  
Using It\^o's lemma,  the L\'evy-driven OU process is such that, for any $t>t_0$
\begin{equation}
\label{eq:solLevyOU}
X_t= e^{-a(t-t_0)}X_{0} + a\int_{t_0}^{t}{b(\theta)e^{-a(t-\theta)}d\theta}  + \sigma \int_{t_0}^{t}{e^{-a(t-\theta)}dY_{c \theta}}.
\end{equation}
and, using (\ref{bra}) and (\ref{eq:solLevyOU}), the integral $\int_{t_0}^{t}{X_{u}du}$ can be reformulated as
\begin{eqnarray}
\label{cond}
\int_{t_0}^{t}{X_{u}du}= \phi(t-t_0)X_{0}+a\int_{t_0}^{t}{b(u)\phi(t-u)du} +\sigma\int_{t_0}^{t}{\phi(t-u)dY_{cu}}.
\end{eqnarray}
Expression (\ref{eq1}) is obtained from (\ref{ZC_price}) and (\ref{cond})  and by using Lemma 3.1 in \citet{Eberlein1999}.
%
\finproof

When $b$ is assumed to be a piecewise-constant function of time as defined by  (\ref{eq:piecewise_const}), the integral in the right hand side of (\ref{eq1}) can be discretized on the time grid $(T_i)_{i=0, \ldots, n}$ which immediately leads to the following proposition.

\begin{pro}
\label{Prop:Levy-OU}
Let $t$ be such that $T_{i-1} < t \le T_i$. In the L{\'e}vy-driven OU model, if $b$ is a step function defined by (\ref{eq:piecewise_const}), then
\begin{equation}
\label{Survival_Probe}
P(t_0,t) = \exp\left( - I(t_0,t,X_0) \right)
\end{equation}
where
\begin{equation}
I(t_0,t,x) := x\phi(t-t_0) + \sum_{k=1}^{i-1}b_k \left(\xi(t-T_{k-1})-\xi(t-T_{k})\right)+b_{i}\xi(t-T_{i-1})  + c \psi(t-t_0)
\end{equation}
and where the functions $\phi$ and $\psi$ are given respectively by (\ref{eq:phi}) and (\ref{eq:psi}) and $\xi$ is defined by
\begin{equation}
\label{eq:xi}
\xi(s):= s -\phi(s).
\end{equation}
\end{pro}

\begin{rem} Note that the function $\phi$ (and thus $\xi$) does not depend on the L\'evy process specification. Moreover, for most L\'evy processes, the integral of the cumulant transform in (\ref{eq:psi}) has no simple closed-form solution but can be easily computed numerically. The reader is referred to \citet{Hainaut2007} for examples of L\'evy processes for which the function $\psi$ defined by (\ref{eq:psi})  admits a closed-form expression.
\end{rem}


Similar analytical expressions are available when the underlying short-rate (or default-intensity) process follows an extended CIR process with deterministic long-term mean parameter $b$.
\begin{lem}\label{cir:affine}
In the extended CIR model (\ref{cir}), the value at time $t_0$ of a generic elementary quantity with maturity $t$ is given by
\begin{equation}
\label{feyn}
P(t_0,t) =\exp\left(-X_0\varphi(t-t_0 ) -a \int_{t_0}^t \varphi(t-u  ) b(u) du\right)
\end{equation}
where $\varphi $ is given by 
\begin{equation}
\label{eqq:xi_CIR}
\varphi(s):=\frac{2(1-e^{-hs})}{h+a + (h-a)e^{-hs}} 
\end{equation}
and $h := \sqrt{a^2 + 2\sigma^2}.$
\end{lem}

\proof
For any maturity date $t$, thanks to the Feynman-Kac formula, the function $\tilde{P}$ defined for any $u$ such that $t_0\leq u \leq t$ by
$$
\tilde{P}(u,x) := \E_{\Q}\left[ \exp\left(-\int_{u}^t{ X_u du}\right) \mid X_u = x\right]
$$
is solution of the following PDE
\begin{equation}
\label{kac}
 \frac{\partial \tilde{P}(u,x)}{\partial u} + a\left( b(u)-x\right)\frac{\partial \tilde{P}(u,t)}{\partial x} +\frac{1}{2}\sigma^2x\frac{\partial^2\tilde{P}(u,t)}{\partial x^2} - \tilde{P}(u,t)x=0,
\end{equation}
with the final condition $\tilde{P}(t, x)=1$, for all $x$. It is straightforward to check that the function $\tilde{P}$ defined by
$$
\tilde{P}(u,x) =\exp\left(-x\varphi(t-u ) -a \int_{u}^t \varphi(t-s  ) b(s) ds\right)
$$
with $\varphi $ given by  (\ref{eqq:xi_CIR}) is solution of PDE (\ref{kac}).
\finproof
Replacing $b$ in (\ref{feyn}) by the piecewise-constant function defined by (\ref{eq:piecewise_const}) yields the following result. 


\begin{pro}
\label{Prop:CIR}
Let $t$ be such that $T_{i-1} < t \le T_i$. In the extended CIR model, if $b$ is a step function defined by (\ref{eq:piecewise_const}), then
\begin{equation}
\label{Survival_Prob}
P(t_0,t) = \exp\left( - I(t_0,t,X_0) \right)
\end{equation}
where
\begin{equation}
I(t_0,t,x) := x\varphi(t-t_0) + \sum_{k=1}^{i-1}b_k \left(\eta(t-T_{k-1})-\eta(t-T_{k})\right) +b_{i} \eta(t-T_{i-1})  
\end{equation}
and where the function $\varphi $ is defined by (\ref{eqq:xi_CIR}) and the function $\eta$ is given by
\begin{equation}
\eta(s):= 2a\left[\frac{s}{h+a} +  \frac{1}{\sigma^2}\log\frac{h+a+(h-a)e^{-hs}}{2h}\right]
\end{equation}
and $h := \sqrt{a^2 + 2\sigma^2}.$
\end{pro}

The previous result can also be found in \citet{Bielecki2014} under a more general form. \citet{Schlogl2000} also consider an extended CIR model with piecewise-constant parameter in order to construct initial yield-curves but prices of zero-coupon bonds  are given in a recursive way in their approach whereas they are expressed in closed-form here.

\begin{rem}
In the case of credit curve construction and in the perspective of curve projection, 
 the positiveness of the default intensity process is guaranteed in the extended CIR model under the Feller's condition or in the L\'evy-driven OU model when choosing a L\'evy-subordinator as L\'evy driver.
\end{rem}

Depending on the chosen term-structure model and given that Assumption \ref{Ass:linear} holds, Proposition \ref{Prop:Levy-OU} or Proposition \ref{Prop:CIR} can be used to compute the present values of instruments selected for the curve construction. Note that, contrary to some L\'evy-driven OU models, no numerical integration is required under the extended CIR specification.

\subsection{Admissible curve construction}
\label{subsec:admissible}

We now explain how to construct admissible curves as described in Definition \ref{Def:admissible}. Recall that the curve is built by matching a set of market quotes $\mathbf{S}=(S_1, \ldots, S_n)$ corresponding to a series of financial products with increasing maturities $\mathbf{T}=(T_1, \ldots, T_n)$.
Under assumption \ref{Ass:linear},  the curve $t \rightarrow P(t_0, t)$ is compatible with the input set  $\mathbf{S}$ if for some payment time grid $(t_1, \ldots,t_{p_1}, \ldots, t_{p_n})$ with $t_{p_i} = T_i$, the column vector  $\mathbf{P} = (P(t_0, t_k))_{k=1, \ldots, p_n}$ is solution of the following rectangular linear system 
\begin{equation}
\label{eq:linear_sys}
\mathbf{A}\cdot \mathbf{P}  = \mathbf{B}
\end{equation}
where
$\mathbf{A}$ is a $n\times p_n$ matrix and $\mathbf{B}$ is a $n\times 1$ matrix. We moreover assume that, for any $i=1, \ldots, n$, the $i$-th line of the previous system corresponds to the market-fit condition of the financial product with maturity $T_i$. Note that matrices  $A$ and $B$ only depend on market quotes $\mathbf{S}$, on standard maturities  $\mathbf{T}$ and on products characteristics. For OIS discounting curve construction, $P = P^D$ and matrices $\mathbf{A}$ and  $\mathbf{B}$  can be easily extracted from system  (\ref{eq:OIS_market_fit}). For credit curve construction based on CDS spreads, $P = Q$ and matrices $\mathbf{A}$  and  $\mathbf{B}$ can be obtained from a discretized version of the system described by (\ref{eq:credit_system}).
As a rectangular system ($n<p_n$),  (\ref{eq:linear_sys}) may admit several solutions\footnote{If maturity dates are strictly increasing, i.e., $T_1<\ldots<T_n$, $\mathbf{A}$ is a full rank matrix (rank $n$) and the solutions of  (\ref{eq:linear_sys}) evolves in a linear space with dimension equal to $n-p_n$.}.\\

We now consider that the curve $t \rightarrow P(t_0,t)$ is given by either Proposition \ref{Prop:Levy-OU} or Proposition \ref{Prop:CIR}. It is straightforward to remark that, for any $i=1, \ldots, n$, when $t$ belongs to the time interval $(t_0, T_i)$, $P(t_0, t)$ only depends on $b_1, \ldots, b_i$. As the $i$-th line of the market-fit system \ref{eq:linear_sys} only involves the curve values at maturity dates smaller than $T_i$,  
solving the previous rectangular system of $\mathbf{P}$ amounts to solving a triangular non-linear system of $\mathbf{b} = (b_1, \ldots, b_n)$ which can be solved iteratively. 
\begin{itemize}
\item  \textbf{Step 1}: Find  $\bar{b}_1$ as the solution of 
\begin{equation}
\label{step1:bootstrap}
\sum_{j=1}^{p_1} A_{1j} P(t_0, t_j;\, b_1)  = B_1
\end{equation}
\item  \textbf{Step 2}: For $k=2, \ldots, n$, assume that $\bar{b}_1, \cdots, \bar{b}_{k-1}$ are known and find $\bar{b}_k$ as  the solution of 
\begin{equation}
\label{step2:bootstrap}
\sum_{j=1}^{p_k} A_{kj} P(t_0, t_j;\, \bar{b}_1, \cdots,\bar{b}_{k-1}, b_k)  = B_k
\end{equation}
\end{itemize}
where $B_k$ denotes the $k$-th element of vector $\mathbf{B}$ and $A_{kj}$ denotes the $(k,j)$-entry of matrix $\mathbf{A}$. 
In most situations, all entries of $A$ have the same sign, so that the left hand side of (\ref{step2:bootstrap}) is a monotonic function of $b_k$. Then, if a solution exists, it is the only one. The previous algorithm is a so-called bootstrap procedure where the resolution of a triangular system of non-linear equations is reduced to successive resolution of univariate equations. Given that the equations are monotonic in the unknown parameter, a numerical solution can be obtained very efficiently at each step by a root-solver.\\

\begin{rem}
Note that, for any $i=1, \ldots,n$, if the implied parameter  $\bar{b}_i$ exists, the latter depends on market quotes $S_1, \ldots, S_i$ and on cash-flow characteristics of the products with maturities $T_1, \ldots, T_i$. In addition,  $\bar{b}_i$ depends on 
the underlying model parameters, that is $\mathbf{p} := (X_0, a, \sigma, c, \mathbf{p}_L)$ for L\'evy-OU models or $\mathbf{p} := (X_0, a, \sigma)$ for the extended CIR model.
\end{rem}

As soon as an implied set of parameter $(\bar{b}_1, \ldots, \bar{b}_n)$ can be found by the previous iterative procedure, the market fit condition is satisfied. However, an admissible curve as described in Definition \ref{Def:admissible} has to fulfill two additional requirements: the curve has to be smooth enough and arbitrage-free. How to be sure that the curve generated with these implied parameters have these two additional features?

\begin{pro}
\label{pro:smith} 
A curve $t \rightarrow P(t_0, t)$ constructed in the previous mean-reverting term-structure models has a derivative which is absolutely continuous. As a consequence, the curve satisfies the smoothness condition described in Definition \ref{def:smooth}.
\end{pro} 

\proof
Let us consider a curve constructed from the L\'evy-OU term-structure model. From equation  \ref{eq1}, the curve $t\rightarrow P(t_0,t)$ is continuous and
its derivative with respect to $t$ is given by
\begin{equation}
\label{deer}
\frac{\partial{P(t_0,t)}}{\partial{t}}=P(t_0,t)\left(-X_{0}e^{-a(t-t_0 )} -a \int_{t_0}^t e^{-a(t-u  )} b(u) du +c\kappa(-\sigma \phi(t-t_0)) \right)
\end{equation}
Therefore, the corresponding instantaneous forward curve is given by
\begin{equation}
\label{deere}
f(t_0,t)=X_{0}e^{-a(t-t_0 )} +a \int_{t_0}^t e^{-a(t-u  )} b(u) du -c\kappa(-\sigma \phi(t-t_0))
\end{equation}
which is an absolutely continuous function of $t$ even if $b$ is a piecewise-constant function of time. As for credit curves, the density function of the underlying default time $t \rightarrow P(t_0,t)f(t_0,t)$ is also absolutely continuous as a product of two absolutely continuous functions.
Given equation \ref{feyn}, the same arguments hold for curves constructed from an extended CIR model where the instantaneous forward rates are given by 
\begin{eqnarray}
\label{eq:f_CIR}
f^{CIR}(t_0, t) = X_{0}\varphi'(t-t_0) + a \int_{t_0}^t \varphi'(t-u)b(u) du
\end{eqnarray}
where $\varphi'$ denotes the derivative of function $\varphi$ defined by (\ref{eqq:xi_CIR}).
\finproof


We proved that a curve constructed from our approach satisfies the smoothness condition. In order to comply with the arbitrage-free requirement, one has to check whether the corresponding instantaneous forward curve is truly positive.  Note that,  given (\ref{deere}) and (\ref{eq:f_CIR}),  instantaneous forward rates have  closed-form expressions in both approaches. Assume that an implied set of mean-reverting parameters $\mathbf{b} = (\bar{b}_1, \ldots, \bar{b}_n)$ have been found using the previous iterative procedure. Let $t$ be a maturity time such that $T_{i-1} \leq t < T_i$. The L\'evy-OU implied forward rate is given by 
\begin{equation}
\label{eq:forward_rates_Levy}
f(t_0, t) = X_0e^{-a(t-t_0)} + a\sum_{k=1}^{i-1}\bar{b}_k\left(\phi(t-T_{k-1}) - \phi(t-T_{k}) \right) + a\bar{b}_i\phi(t-T_{i-1})  -c\kappa(-\sigma \phi(t-t_0))
\end{equation}
and the implied CIR forward rate is given by 
\begin{equation}
\label{eq:forward_rates_CIR}
f^{CIR}(t_0, t) = X_0\varphi'(t-t_0) + a\sum_{k=1}^{i-1}\bar{b}_k\left(\varphi(t-T_{k-1}) - \varphi(t-T_{k})\right) + a\bar{b}_i\varphi(t-T_{i-1}).
\end{equation}
A naive method would consist in checking the positivity of forward rates for every time point between $t_0$ and $T_n$.
The next proposition shows that for L\'evy-OU term-structures, the positivity of constructed forward rates can be checked in a very efficient way. 
\begin{pro}
\label{pro:arbitrage-free_LevyOU}
 Let $t\rightarrow P(t_0, t)$ be a curve constructed from a L\'evy-OU term-structure model  and assume that the vector $\mathbf{b} = (\bar{b}_1, \ldots, \bar{b}_n)$ of implied mean-reverting levels exists.  
  The forward curve $t\rightarrow f(t_0, t)$ is then given by (\ref{eq:forward_rates_Levy}). 
 Assume that  $\mathbf{p}:=(X_0, a, \sigma, c, \mathbf{p}_L)$ is a vector of positive parameters and that the derivative of the L\'evy cumulant  $\kappa'$ exists and is strictly monotonic on $(-\infty, 0)$. The constructed curve is arbitrage-free on the time interval $(t_0, T_n)$  if and only if, for any $i=1, \ldots, n$, $f(t_0, T_{i}) > 0$ and one of the following condition holds:
 \begin{itemize}
 \item $\frac{\partial f}{\partial t}(t_0, T_{i-1}) \frac{\partial f}{\partial t}(t_0, T_{i})\geq 0$, 
 \item$\frac{\partial f}{\partial t}(t_0, T_{i-1}) \frac{\partial f}{\partial t}(t_0, T_{i})<0$ and $f(t_0, t_i) > 0$ where $t_i$ is such that $\frac{\partial f}{\partial t}(t_0, t_i) = 0$,
 \end{itemize}
 where we recall that $T_0:=t_0$.
 \end{pro}

\proof 
The curve is arbitrage-free if $P(t_0, t)$ is a nonincreasing function of $t$.
From equations (\ref{eq1}) and (\ref{deer}), this is truly the case if $f(t_0, t)$ given by (\ref{eq:forward_rates_Levy}) is positive for any $t$ in the interval $(t_0, T_n)$. Note that, $f(t_0, T_0) = f(t_0, t_0) = X_0>0$ and for any $i=1, \ldots, n$ and any $t$ such that $T_{i-1} \leq t < T_i$, the forward rate $f(t_0, t)$ can be expressed as a function of $f(t_0, T_{i-1})$ and $\bar{b}_i$:

\begin{equation}
\label{eq:forward_rates_bi}
\begin{split}
f(t_0, t) =  \bar{b}_i &+ \left[f(t_0, T_{i-1}) +  c\kappa(-\sigma \phi(T_{i-1}-t_0)) - \bar{b}_i \right]e^{-a(t-T_{i-1})}\\ 
&-  c\kappa\left(-\frac{\sigma}{a}\left( 1- e^{-a(T_{i-1}-t_0)}e^{-a(t-T_{i-1})} \right)\right)
\end{split}
\end{equation} 
Let $g_i$ be the function defined on $(T_{i-1}, T_i)$ and such that $g_i(t) = \exp(-a(t-T_{i-1}))$. On $(T_{i-1}, T_i)$, the function $t\rightarrow f(t_0, t)$ is such that $f(t_0, t) = K_i(g_i(t))$ where $K_i$ is defined on $(g_i(T_{i}), 1)$ by
\begin{equation}
\label{eq:Ki}
K_i(x) =  \bar{b}_i + \left[f(t_0, T_{i-1}) +  c\kappa(-\sigma \phi(T_{i-1}-t_0)) - \bar{b}_i \right]x
-  c\kappa\left(-\frac{\sigma}{a}\left( 1- e^{-a(T_{i-1}-t_0)}x \right)\right).
\end{equation} 
Note that $f(t_0, t)$ is strictly positive on $(T_{i-1}, T_i)$ if and only if  $K_i$ is strictly positive on the interval $(g_i(T_{i}), 1)$. As $f(t_0, T_{i-1})>0$ and   $f(t_0, T_{i})>0$, then $K_i$ is also strictly positive at the extreme points $g_i(T_{i})$ and $1$. In addition, if $\kappa$ is differentiable, the first derivative of $K_i$ is given by 
\begin{equation}
\label{eq:Ki_diff}
K_i'(x) = f(t_0, T_{i-1}) +  c\kappa(-\sigma \phi(T_{i-1}-t_0)) - \bar{b}_i 
-  \frac{c\sigma}{a}e^{-a(T_{i-1}-t_0)}\kappa'\left(-\frac{\sigma}{a}\left( 1- e^{-a(T_{i-1}-t_0)}x \right)\right).
\end{equation} 
As $\kappa'$ is assumed to be a strictly monotonic  function on $(-\infty, 0)$, $K_i'$ is also a strictly monotonic function on the interval $(g_i(T_{i}), 1)$ as an affine transformation of a composition of $\kappa'$ with an affine function of $x$. Note that $\frac{\partial f}{\partial t}(t_0, t) = K_i'(g_i(t))\cdot g_i'(t)$ where $g_i'(t)<0$ for all $t$ in $[T_{i-1}, T_i]$. Then, let us deal with the following three possible situations:
\begin{itemize}
\item If $\frac{\partial f}{\partial t}(t_0, T_{i-1}) \frac{\partial f}{\partial t}(t_0, T_{i})> 0$, then $K_i'(g_i(T_{i})) K_i'(1) > 0$. As a result,  $K_i'(g_i(T_{i}))$ and $K_i'(1)$ have the same sign and, since $K_i'$ is a strictly monotonic function, $K_i'$ cannot cross the $x$-axis. Consequently, $K_i$ is a strictly monotonic function on $(g_i(T_{i}), 1)$. As $K_i$ is strictly positive at the extreme points of $(g_i(T_{i}), 1)$, it remains positive on this interval. 
\item If $\frac{\partial f}{\partial t}(t_0, T_{i-1}) \frac{\partial f}{\partial t}(t_0, T_{i})= 0$, then,  $K_i'(g_i(T_{i}))=0$ or $K_i'(1)=0$ but
 the two previous quantities cannot be equal to zero simultaneously since  $K_i'$ is strictly monotonic. Then, $K_i'$ is either positive or negative  on $(g_i(T_{i}), 1)$ and we can conclude as in the preceding case.
\item If $\frac{\partial f}{\partial t}(t_0, T_{i-1}) \frac{\partial f}{\partial t}(t_0, T_{i})< 0$, then $K_i'(g_i(T_{i})) K_i'(1) < 0$. As a result,  $K_i'(g_i(T_{i}))$ and $K_i'(1)$ have opposite signs and, since $K_i'$ is a strictly monotonic function, $K_i'$ crosses the $x$-axis once on $(g_i(T_{i}), 1)$. Let  $x_i$ be such that $K_i'(x_i)=0$. Then $K_i$ admits a unique extremum at $x_i$. Since $K_i$ is either successively increasing and decreasing or successively decreasing and increasing and as $K_i$ is strictly positive at the extreme points of $(g_i(T_{i}), 1)$, $K_i$  is positive on $(g_i(T_{i}), 1)$ if and only if $K_i(x_i)>0$. This is equivalent to impose that $f(t_0, t_i) > 0$ where $t_i = g_i(x_i)$.

\end{itemize}

\finproof 

Note that, if $Y$ is a L\'evy-subordinator, its cumulant function has the following form (see, e.g., Theorem 1.3.15 in \citet{Applebaum2009}) 
\begin{equation}
\label{eq:cumulant_Levy_subordinator}
\kappa(\theta)=\alpha \theta +\int_{0}^{\infty}\left(e^{\theta y} -1 \right) \rho(dy)
\end{equation}
where $\alpha$ is a positive parameter and $\rho$ is the L\'evy measure of $Y$. Then if $\rho$ has a finite mean,  $\kappa$ is differentiable on $(-\infty, 0)$ since $\frac{\partial }{\partial \theta} \left(e^{\theta y} -1 \right)= ye^{\theta y} \leq y$ for $\theta$ in  $(-\infty, 0)$ and $\int_{0}^{\infty}y \rho(dy) < \infty$. Consequently, the derivative of $\kappa$ on $(-\infty, 0)$  is given by
\begin{equation}
\label{eq:diff_cumulant_Levy_subordinator}
\kappa'(\theta)=\alpha +\int_{0}^{\infty}ye^{\theta y} \rho(dy).
\end{equation}
Then, if the L\'evy-OU term-structure model is driven by a L\'evy subordinator, its cumulant function $\kappa$ is differentiable on $(-\infty, 0)$ as soon as the mean of the underlying L\'evy measure is finite and, from (\ref{eq:diff_cumulant_Levy_subordinator}),  its derivative $\kappa'$ is a strictly increasing function on $(-\infty, 0)$. In the Vasicek term-structure model, the driver is a Brownian motion, then from Table \ref{Table:Levy}, $\kappa'(\theta) = \theta$ and $\kappa'$ is obviously an increasing function. The other examples in Table \ref{Table:Levy} corresponds to L\'evy subordinators and one can check  for these examples that $\kappa'$ is indeed an increasing function. 


\begin{pro}
\label{pro:arbitrage-free_CIR}
 Let $t\rightarrow P(t_0, t)$ be a curve constructed from an extended CIR term-structure model  and assume that the vector $\mathbf{b} = (\bar{b}_1, \ldots, \bar{b}_n)$ of implied mean-reverting levels exists. The forward curve $t\rightarrow f^{CIR}(t_0, t)$ is then given by (\ref{eq:forward_rates_CIR}). The constructed curve is arbitrage-free if $\mathbf{p} := (X_0, a, \sigma)$ is a vector of positive parameters and if,  for any  $i=1, \cdots,n$, the implied $\bar{b}_i$ is positive.
\end{pro}

\proof 
The result follows from equation \ref{eq:forward_rates_CIR} and the fact that $\varphi$ is an increasing function.
\finproof 

\begin{rem}
Note that a weaker condition can be found on the implied $\bar{b}_i$'s to guarantee the positivity of CIR-implied forward rates. However, under this condition, mean-reverting levels could be negative, which is incompatible with a well-defined square-root model.
\end{rem}

Given a set of market quotes $(S_1,\ldots, S_n)$ and a corresponding set of standard maturities $(T_1, \ldots, T_n)$, the curve $t\rightarrow P(t_0, t;\, \bar{b}_1, \ldots, \bar{b}_n)$ is then admissible on $(t_0, T_n)$ if the implied parameters $\bar{b}_1, \ldots, \bar{b}_n$  exist and fulfill the assumptions of Proposition \ref{pro:arbitrage-free_LevyOU} for L\'evy-driven OU models or the assumptions of Proposition \ref{pro:arbitrage-free_CIR} for CIR models. Note that the two previous proposition can be used within the iterative bootstrap algorithm. Indeed, when seeking for the implied mean-reverting levels, the numerical procedure can be stopped as soon as one of the no-arbitrage condition is not satisfied.

\subsection{Numerical illustration}

Let us now present some interest-rate and credit term structures constructed using the previous approaches.\\  


\noindent \textbf{Construction of admissible OIS discounting curves}\\

We first consider the construction of OIS discounting curves based on OIS market quotes as observed in May, 31st 2013 and given in Table \ref{table:OIS_data}. In this example, the L\'evy-driven OU short-rate model (\ref{bra}) is used as generator of admissible curves. 
The L\'evy driver is chosen to be a Gamma subordinator with a cumulant function defined as in Table \ref{Table:Levy} and with parameter $\lambda=200$. This choice of $\lambda$ corresponds to a Gamma subordinator with an annual mean jump size of $50$ bps. Note that, if the short-rate process is given by (\ref{bra}), the parameter $c$ corresponds to the number of jumps that the short-rate is expected to do in a one year period.
In order to illustrate the diversity of admissible discounting curves, different values of the jump frequency parameter $c$ are considered. The starting point $X_0$ of the short rate process is fixed at $0.063\%$ which corresponds to the May, 31st 2013 rate of the OIS with maturity 1 month. The parameters $a$ and $\sigma$ are such that $a=0.01$ and $\sigma=1$. 
For each considered value of $c$, the mean-reverting parameters $b_i$'s are bootstrapped from OIS swap rates using the procedure described in Subsection \ref{subsec:admissible}. Proposition \ref{Prop:Levy-OU} is used to compute discount factors in this approach. Figure \ref{Lev} displays the set of discount factor curves obtained by repeating the construction process for each value of $c$ in the set $\left\{1,10,20,\ldots,100 \right\}$. Each of these values leads to an admissible curve. Figure \ref{Lev2} represents the corresponding set of (continuously-compounded) discount rates and instantaneous forward curves (upper set of curves at low maturities). In  Figure \ref{Lev} and  Figure \ref{Lev2}, the black segments corresponds to arbitrage-free bounds at time-to-maturities $15y$, $20y$, $30y$ and $40y$. These bounds have been computed using Proposition \ref{pro:OIS_bounds_model_free_sharp}. As expected, the values taken by the displayed curves at these maturities belong to the no-arbitrage bounds.\\




\begin{figure}[h]
\begin{center}
\includegraphics[height = 6cm, width = 12cm]{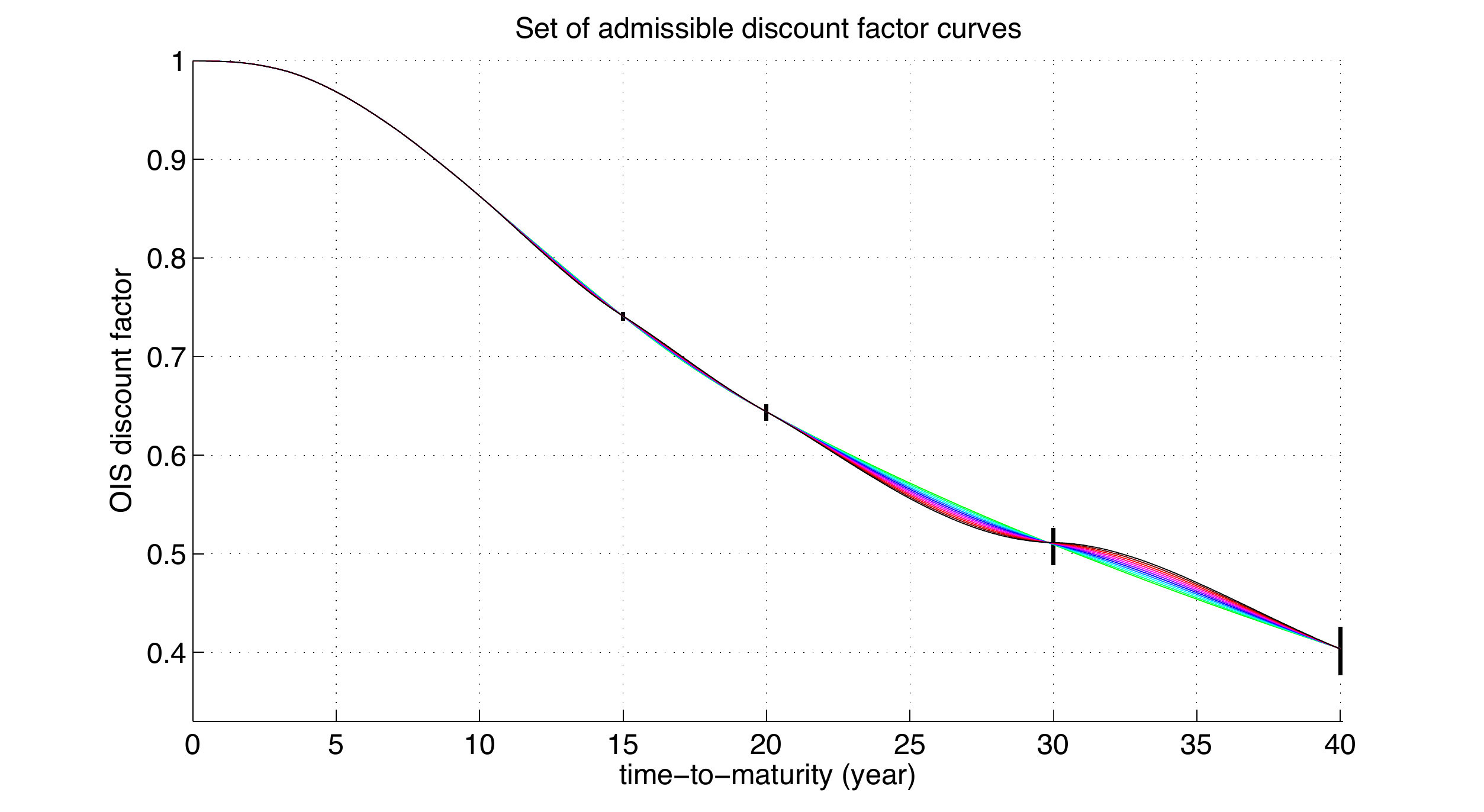}
\caption{ OIS discount curves computed in a L\'evy OU model as of May, 31st 2013}
\label{Lev}
\end{center}
\end{figure}

\begin{figure}[h]
\begin{center}
\includegraphics[height = 6cm, width = 12cm]{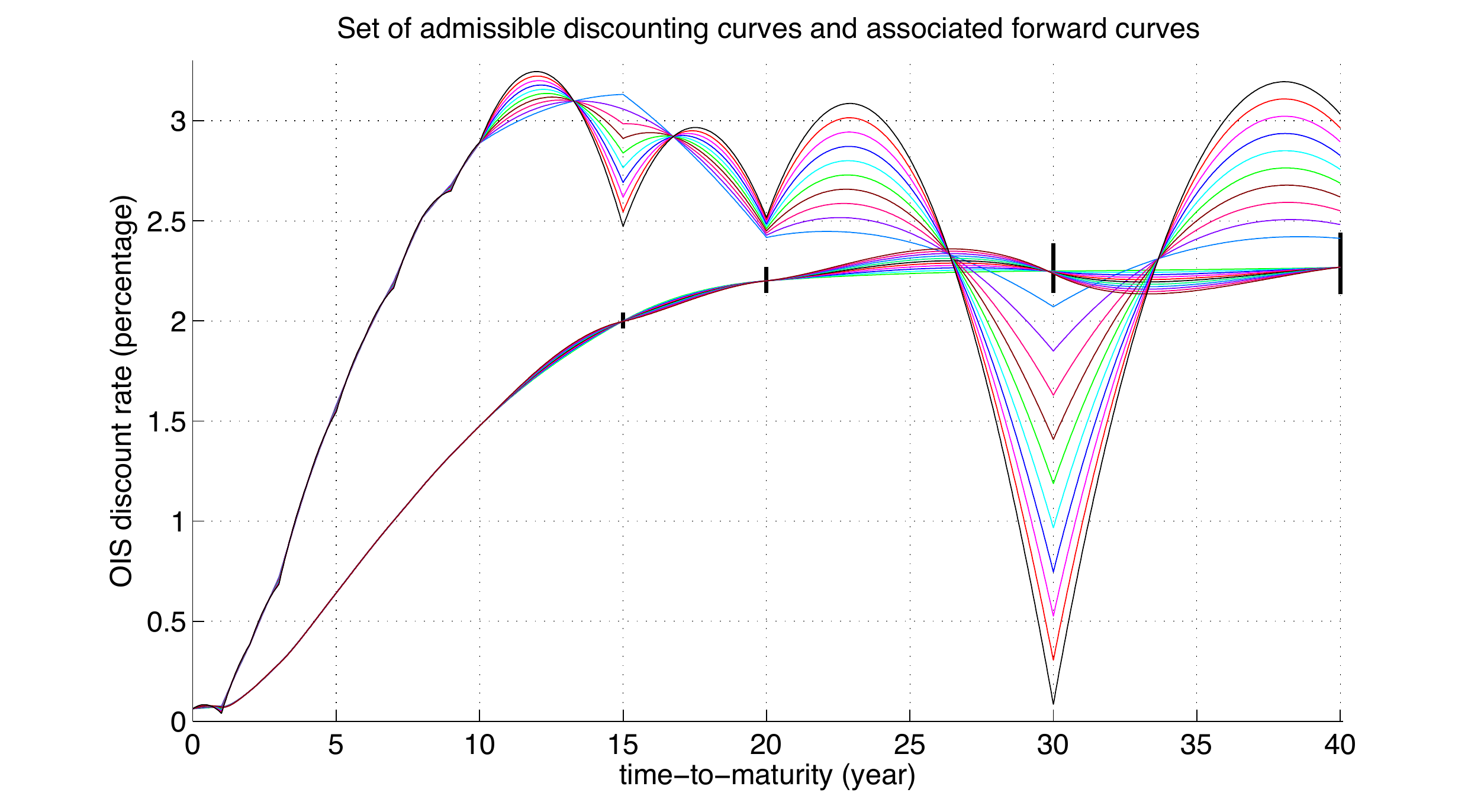}
\caption{ OIS discount rate and associated forward curves computed in a L\'evy OU model as of May, 31st 2013}
\label{Lev2}
\end{center}
\end{figure}

\newpage

As can be seen in Figure \ref{Lev} and Figure \ref{Lev2}, playing with parameter $c$ has little impact on the diversity of values taken by these curves for maturities lower than 10y. This is due to the fact that discount factors are known without uncertainty for these maturities. However, the effect of $c$ on curve diversity is significant as maturities become larger than 10y. For discount factor curves, this effect is exacerbated for the two last maturity periods $(20y, 30y)$ and $(30y, 40y)$.
%
Note that, for maturities greater than 10y, the variability of forward curves (Figure \ref{Lev2}) is much higher than the variability of the associated spot rate curves (Figure \ref{Lev2}) and discount curves (Figure \ref{Lev}). The distance between two forward curves can be close to two points of percentage, as can be seen for instance at maturity $30y$. This suggests that, given a family of admissible construction methods, the range of values taken by the resulting forward curves can be significantly larger than the range of values taken by associated spot rate curves.
It means that the uncertainty embedded in the process of curve construction is amplified for forward curves.\\


As a matter of comparison,  we now consider the extended CIR model (\ref{cir}) as generator of discounting curves. The input OIS data set is the same as in the previous example. The underlying parameters of the CIR short-rate process are chosen such that $X_0 = 0.063\%$ and $a=\sigma=1$.
Contrary to the previous example, the admissible curves are not generated here by playing on some extra free parameters. We instead choose to include additional fit constraints in the calibration process, while preserving the admissible nature of the curves. In other words, the generated curves have been forced to take some pre-specified values at some pre-specified maturity dates. These pre-specified points have been chosen consistently with the no-arbitrage bounds in such a way that the generated curves are admissible and take values close to the upper or lower no-arbitrage bounds at standard maturities.

\begin{figure}[h]
\begin{center}
\includegraphics[scale = 0.5]{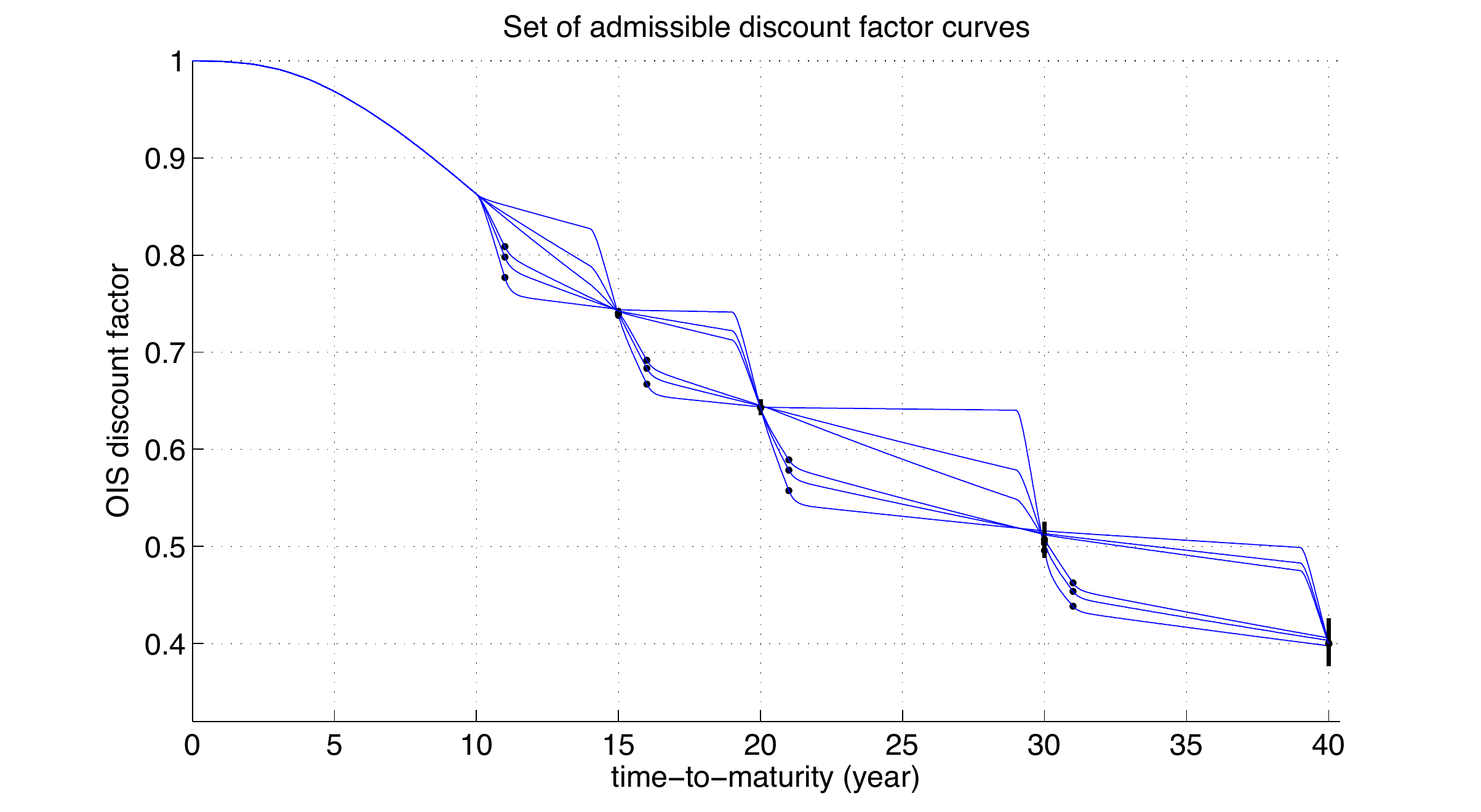}
\caption{ OIS discount factor curves computed from swap rates as of May, 31st 2013}
\label{CIR_1}
\end{center}
\end{figure}

\begin{figure}[h]
\begin{center}
\includegraphics[scale = 0.5]{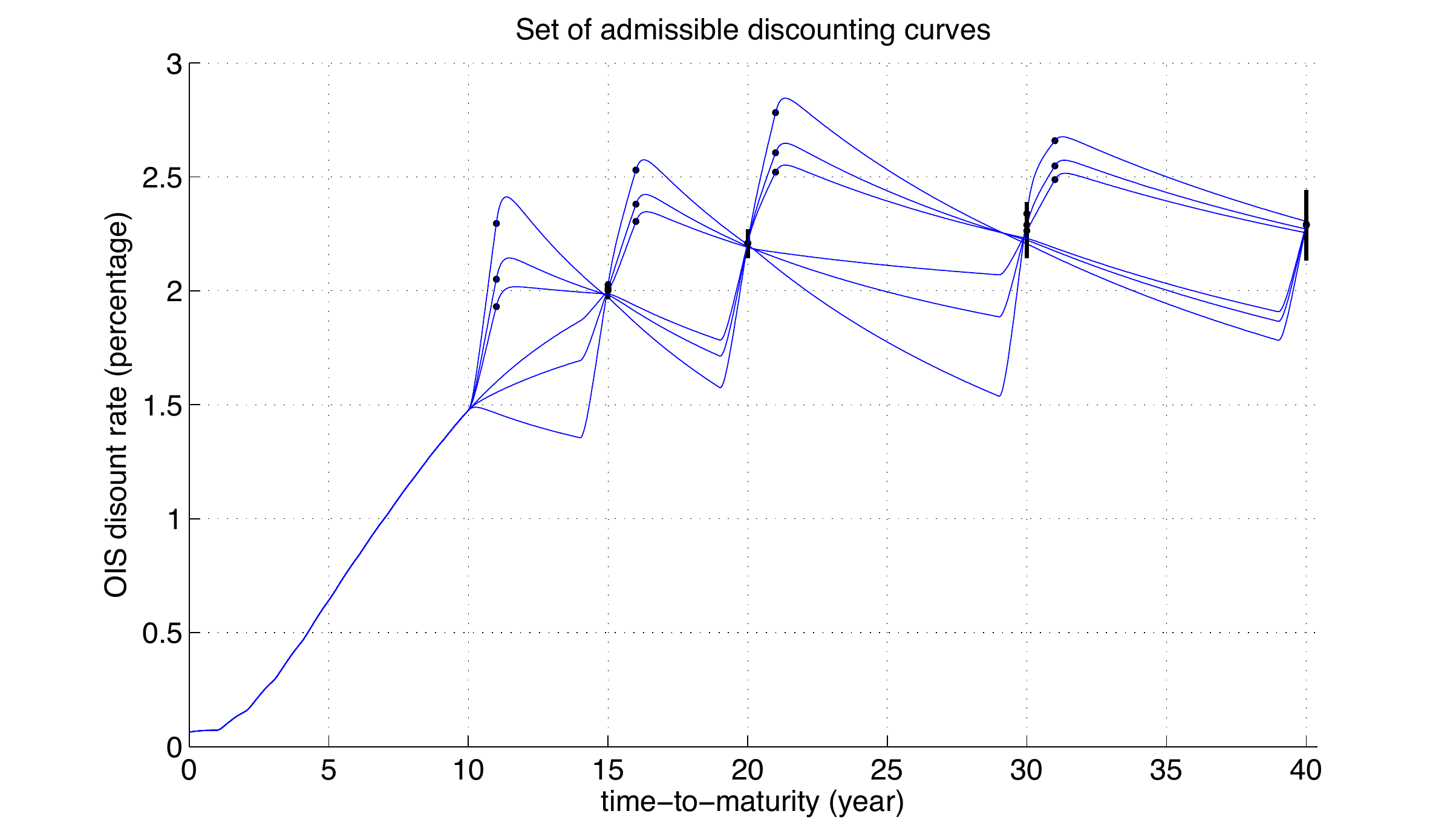}
\caption{OIS discount rate curves computed from swap rates as of May, 31st 2013}
\label{CIR_2}
\end{center}
\end{figure}

 Each discount factor curve plotted on Figure \ref{CIR_1} corresponds to a particular set of pre-specified values that can be identified by a set of black dots. Figure \ref{CIR_2} represents the corresponding set of admissible discount rate curves. The no-arbitrage bounds (which are exactly the same as in Figure \ref{Lev} and Figure \ref{Lev2}) are represented by black segments at standard maturities.\\

As can be seen in Figure \ref{CIR_1} and \ref{CIR_2}, all generated curves behave similarly for maturities lower than 10y. For maturities larger than 10y, the displayed set of admissible curves can take strikingly different values, especially at non-standard maturities. Note that, thanks to the convex nature of the set of admissible curves (see Proposition \ref{Pro:convex}), any point between two admissible curves is reached by an admissible curve. As can be observed in Figure \ref{CIR_2}, the range of admissible discount rates are nearly equal to one point of percentage for some maturities. 
\\

\noindent \textbf{Construction of admissible survival curves}\\

As stressed in the previous sections, our approach can also be used to construct survival curves or default distribution functions from a series of quoted CDS spreads. Each spread represents the cost of protection associated with the same underlying debt issuer but for different protection maturity.
In this example, we consider CDS spreads of AIG for maturities $3y$, $5y$, $7y$ and $10y$ as observed in December 17, 2007 and given in Table \ref{CDS_data}.  The chosen curve generator is the extended  CIR default intensity model (\ref{cir}) where $a=\sigma=1$. As explained in 
Subsection \ref{subsec:admissible}, the piecewise-constant mean-reverting level $b_i$'s are bootstraped from AIG  market spreads. 
Figure  \ref{Cir_pannel} displays the set of admissible survival curves obtained by repeating the construction process for each value of $X_0$ such that $100X_0$ is in the set $\{0.01, 0.25, 0.49,$ $0.73, 0.97,$ $1.21, 1.45, 1.69, 1.94, 2.18, 2.42\}$. All the generated curves are admissible. The arbitrage-free bounds computed from Proposition  \ref{pro:CDS_survi_model_free} are represented by black segments at  standard maturities $3y$, $5y$, $7y$ and $10y$. Note that the survival curves are consistent with the no-arbitrage bounds.\\




\begin{figure}[h]
\begin{center}
\includegraphics[scale=0.5]{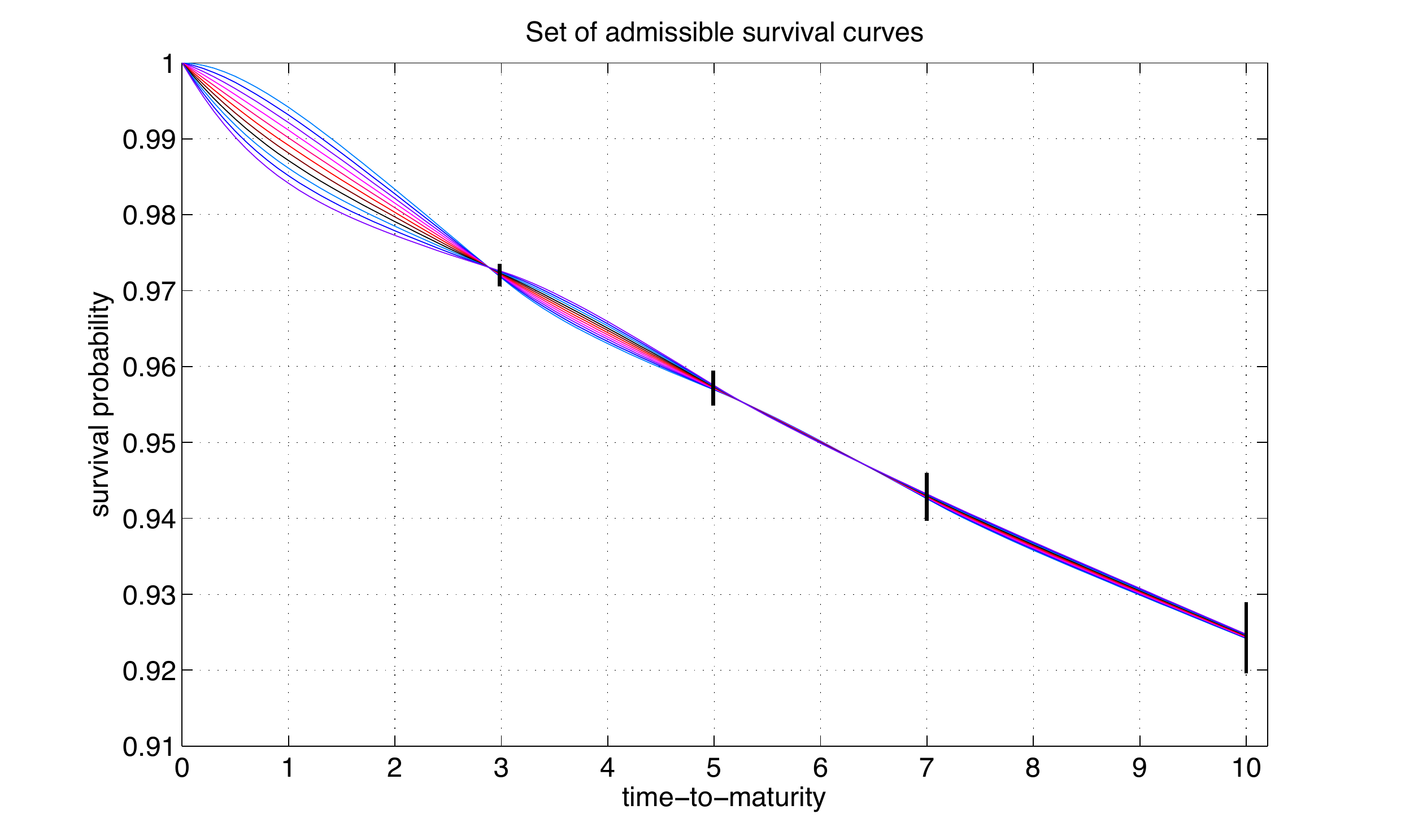}
\caption{Survival curves computed from CDS spreads of AIG as of December 17, 2007. The curves have been computed with $R=40\%$ and a discounting curve such that $P^D(t_0, t) = \exp(-3\% (t-t_0))$.}
\label{Cir_pannel}
\end{center}
\end{figure}

As we can observe, the initial default intensity $X_0$ has a significant impact on survival probabilities for time-to-maturities nearby zero, in particular for the two first time periods $(0, 3y)$ and $(3y, 5y)$. The effect of $X_0$ is less significant for maturities larger than 5y. We can also note that the range of admissible values seem to be larger for maturity point that far from standard maturities. This suggests that a proper way to reduce the uncertainty in the curve construction process could be to enhance market liquidity at the middle points of any two consecutive standard maturities.

\section{Conclusion}
\label{sec:conclusion}

In this paper, we propose a methodology that allows to estimate the diversity of term-structure functions with some admissible features: arbitrage-freeness, market-consistency and a minimum degree of smoothness. We first show how to compute model-free bounds at standard maturities in the class of arbitrage-free and market-consistent term-structure functions. As for OIS discount curves, the proposed bounds are sharp and can be used to detect arbitrage opportunities that could be  hidden in the input market dataset. Similar bounds can be obtained for CDS-implied survival curves. This framework can easily be adapted to bond term-structures. When an additional minimum smoothness condition is required, dynamic term-structure models with a mean-reversion effect are appropriate to generate admissible curves. We show that the diversity of admissible curves can be appreciated in difference situations (OIS discount curve and CDS curve construction) and within different approaches (L\'evy-driven OU or extended CIR model) by playing with some extra unfitted parameters. In addition, as the set of admissible curves is convex, any values between two admissible curves is reached by an admissible curve. The numerical results suggest that, for both OIS discounting curves and CDS survival curves, the operational task of building term-structures may be associated with a significant degree of uncertainty. This kind of model risk should be, in our view, considered with more attention. 
Measuring the impact of curve diversity on valuation and hedging of financial products is a next step which is part of an ongoing research project.
 Another perspective should be to extend the proposed framework to a multi-curve interest-rate environment, where several curves (with possibly different tenors and different currencies) have to be constructed in a joint consistent process.





\bibliographystyle{abbrvnat.bst}

\bibliography{biblio}

\end{document}